\documentclass[a4paper]{article}

\pdfoutput=1
\usepackage{indentfirst}
\usepackage{amsmath}
\usepackage[pdftex]{graphicx}
\usepackage{longtable}
\usepackage{setspace}
\usepackage{threeparttable}
\title{Mid-infrared spectroscopy of zodiacal emission with AKARI/IRC}
\author{Aoi \textsc{Takahashi} $^{1,2,3,4}$, Takafumi \textsc{Ootsubo} $^{3}$, Hideo \textsc{Matsuhara} $^{3,4}$, \\
Itsuki \textsc{Sakon} $^{5}$, Fumihiko \textsc{Usui} $^{6}$, Hiroki \textsc{Chihara} $^{7}$}
\date{August 4 2019}

\begin{document}

\maketitle

{\small {\sl
\begin{enumerate}
\item Astrobiology center, 2-21-1 Osawa, Mitaka, 181-8588, Japan
\item National Astronomical Observatory of Japan, 2-21-1 Osawa, Mitaka, 181-8588, Japan
\item Institute of Space and Astronautical Science, Japan Aerospace Exploration Agency, 3-1-1 Yoshinodai, Chuo-ku, Sagamihara 252-5210, Japan
\item Department of Space and Astronautical Science, The Graduate University for Advanced Studies (SOKENDAI), Shonan Village, Hayama 240-0193, Japan
\item Graduate School of Science, The University of Tokyo, 7-3-1 Hongo, Bunkyo-ku, Tokyo 113-0033, Japan
\item Center for Planetary Science, Graduate School of Science, Kobe University, 7-1-48 Minatojima-Minamimachi, Chuo-Ku, Kobe 650-0047, Japan
\item Department of General Education, Osaka Sangyo University, 3-1-1, Nakagaito, Daito, Osaka 560-0043, Japan
\end{enumerate}
* Corresponding author: aoi.takahashi@nao.ac.jp
}}\\

\begin{abstract}

Interplanetary dust (IPD) is thought to be recently supplied from asteroids and comets. 
Grain properties of the IPD can give us the information about the environment in the proto-solar system, and can be traced from the shapes of silicate features around 10~$\mu$m seen in the zodiacal emission spectra. 
We analyzed mid-infrared slit-spectroscopic data of the zodiacal emission in various sky directions obtained with the Infrared Camera on board the Japanese AKARI satellite. 
After we subtracted the contamination due to instrumental artifacts, we have successfully obtained high signal-to-noise spectra and have determined detailed shapes of excess emission features in the 9--12~$\mu$m range in all the sky directions. 
According to a comparison between the feature shapes averaged over all directions and the absorption coefficients of candidate minerals, the IPD was found to typically include small silicate crystals, especially enstatite grains. 
We also found the variations in the feature shapes and the related grain properties among the different sky directions. 
From investigations of the correlation between feature shapes and the brightness contributions from dust bands, the IPD in dust bands seems to have the size frequency distribution biased toward large grains and show the indication of hydrated minerals. 
The spectra at higher ecliptic latitude showed a stronger excess, which indicates an increase in the fraction of small grains included in the line of sight at higher ecliptic latitudes.  
If we focus on the dependence of detailed feature shapes on ecliptic latitudes, the IPD at higher ecliptic latitudes was found to have a lower olivine/(olivine+pyroxene) ratio for small amorphous grains. 
The variation of the mineral composition of the IPD in different sky directions may imply different properties of the IPD from different types of parent bodies, because the spatial distribution of the IPD depends on the type of the parent body. 

\end{abstract}

\clearpage

\section{Introduction}
\label{intro}

Asteroids and comets are primordial planetesimals, which have been formed in the proto-solar disk by gathering surrounding dust. 
Their constituents will surely retain the information about the environment in the proto-solar system. 
However, the opportunity to observe comets is limited, and furthermore, nm-scaled structures in regolith on asteroids surfaces have well-experienced two types of space weathering~\cite{Vernazza_2009, Noguchi_2011}: micrometeoroid bombardment, on a timescale of 10$^{8}$--10$^{9}$ years~\cite{Sasaki_2001} and solar-wind irradiation, on a timescale of 10$^{4}$--10$^{6}$ years~\cite{Strazzulla_2005} in case of the near-Earth asteroids. 

On the other hand, interplanetary dust (IPD) can be always observed and tell us the information on the proto-solar phase. 
The IPD is the population of mineral dust grains with a size frequency distribution expressed by a smoothly broken power-law with negative indices in the size range from sub-$\mu$m to millimeters (\cite{Grun_1985}, \cite{Jehn_2000}). 
It distributes in the interplanetary space globally and diffusely, and is thought to be supplied continuously from asteroids and comets. 
The IPD formed by the collisional cascades of asteroids~\cite{Sykes_1986, Nesvorny_2003, Nesvorny_2006} is mainly supplied from the internal of asteroids, which has not seriously been affected by the space weathering. 
The IPD supplied by comets disruption~\cite{Nesvorny_2010} is also from the internal of comets, and the IPD liberated by comets sublimation~\cite{Liou_1995, Poppe_2016} is supplied from the comets surface, which is refreshed every time they approach to the perihelion. 
Once the IPD has been liberated from the parent bodies, it accretes into the Sun along a spiral orbit due to the Poynting-Robertson drag. 
The accretion timescale is typically 10$^{4}$--10$^{5}$ years for 1~$\mu$m-sized grains, although it increases proportionally to the grain size~\cite{Burns_1979, Wyatt_1999}. 
It is comparable to or shorter than the timescale of space weathering for internal structures of the IPD, which is assumed to be similar to that for nm-scaled structures in asteroids regolith. 
Thus, at least small grains in the IPD can be thought to accrete into the Sun before their properties are altered well by space weathering. 
This means the IPD we currently observe, especially about small grains, has not significantly been affected by the space weathering, and the grain properties such as compositions and crystal morphologies reflect dust properties in the formation regions of the parent planetesimals, like asteroids or comets. 
Investigations of the grain properties of the IPD are, therefore, important for studies of the formation process of the planetary system. 

The IPD absorbs sunlight in the ultraviolet--near-infrared and re-emits the absorbed energy as thermal emission in the mid-infrared. 
We observe it as zodiacal emission, which is the dominant source of diffuse mid-infrared radiation all over the sky. 
Some spectral features have been detected in past spectroscopic observations of the zodiacal emission.
The mid-infrared camera on board the Infrared Space Observatory (ISOCAM) measured sky spectra over the wavelength range of 5--16~$\mu$m and reported excess emission in the 9--11~$\mu$m range, with an amplitude of 6\% of the continuum~\cite{Reach_2003}. However, Leinert~et~al.~(2002)~\cite{Leinert_2002} concluded that the zodiacal emission spectra obtained with the spectrophotometric sub-instrument of the ISO photometer, ISOPHOT, are smooth and featureless. 
The Mid-infrared Spectrometer on board the Infrared Telescope in Space (IRTS), the first Japanese infrared telescope, obtained zodiacal emission spectra in the 4.5--11.7~$\mu$m range and reported a 9--12~$\mu$m excess~\cite{Ootsubo_1998}. 
These emission features originate from the Si-O vibration modes in small silicate grains of the IPD. 
The shapes of these features depend on the grain properties of the IPD, such as the mineral and chemical composition, crystallinity, and crystal morphology, for example. 
These feature shapes can thus be used as tracers for investigating the properties of small IPD grains and can help us to understand the environment in the proto-solar system. 
Past observations of the zodiacal emission, however, have not been sensitive enough to determine detailed shapes of the excess emission features. 

In the present paper, we derive the zodiacal emission spectra with high signal-to-noise ratios from 74 observations toward various sky directions, and we present the detailed shapes of the silicate features for the first time. 
The following two sections explain the observations and the data-reduction process. 
In section 4, we discuss the average feature shape and the variations in different sky directions, and we demonstrate a possible link to the grain properties of the IPD. 
In section 5, we present the implications of these results, and we summarize our conclusions in the final section.

\section{Observations}

AKARI is a Japan-lead infrared astronomy satellite~\cite{Murakami_2007}. It carries a telescope with an effective aperture diameter of 68.5 cm and two focal plane instruments. One of them is the Infrared Camera (IRC; \cite{Onaka_2007}). 
The IRC has three channels with different wavelength coverages: the NIR channel for 2--5~$\mu$m, the MIR-S channel for 5--13~$\mu$m, and the MIR-L channel for 12--27~$\mu$m (although spectroscopy in the latter is limited to the region longer than 17.5~$\mu$m, owing to an instrumental problem). 
All three channels operate simultaneously. 
Each channel is equipped with a slit aperture for spectroscopic observations of diffuse sources. 
The filter wheel in each channel includes a blank window as a shutter, and it enables accurate subtraction of dark-current images of the detector array, which is necessary for high-sensitive observations of diffuse sources. 

\begin{figure}[h!]
\begin{center}
     \includegraphics[width=0.9\linewidth]{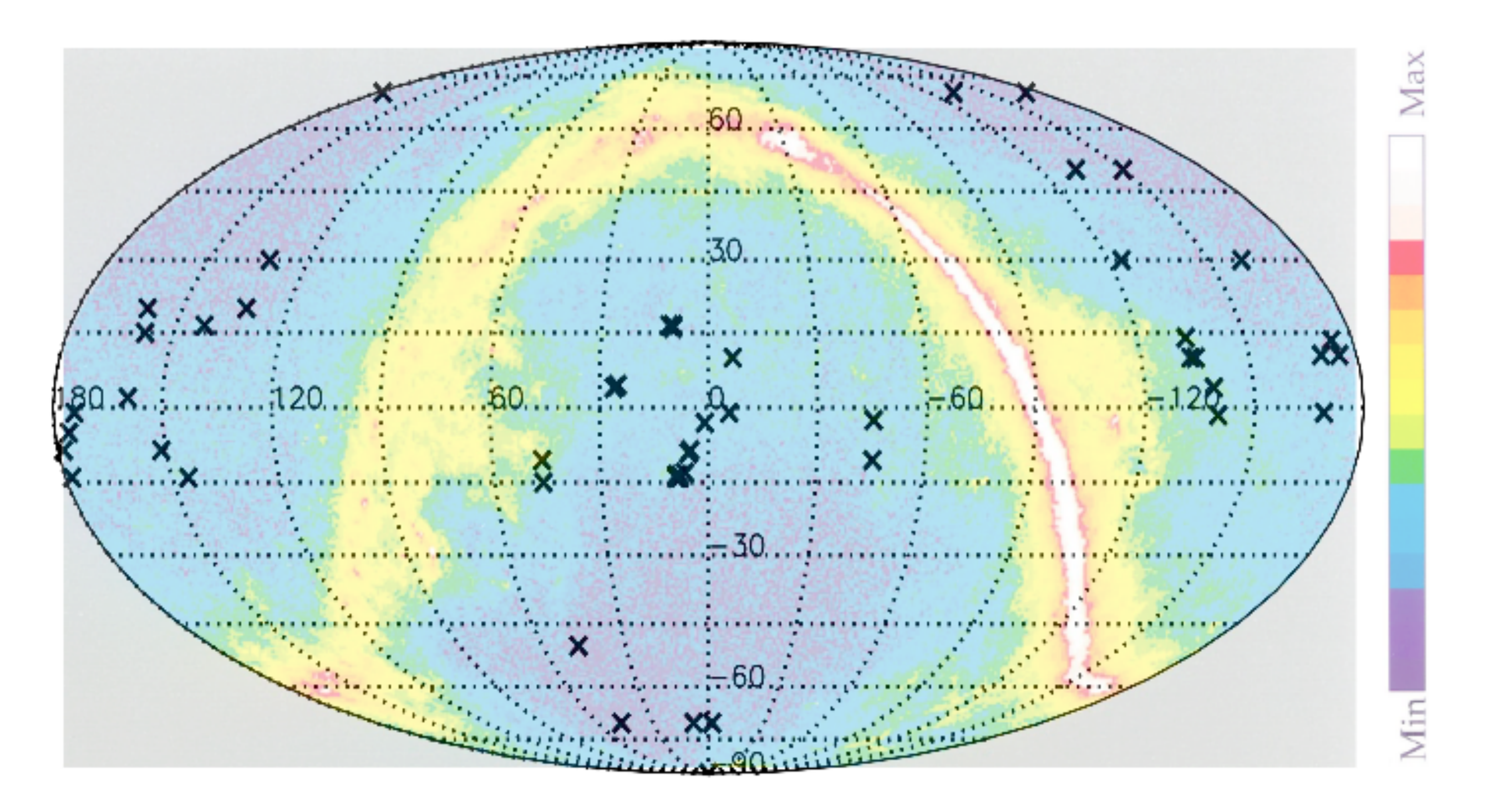}
\end{center}
\caption{Map of the pointing directions we observed. This is a Mollweide projection in geocentric ecliptic coordinates. The background image is the brightness map at 140~$\mu$m obtained with COBE/DIRBE~\cite{Kelsall_1998}, which is dominated by thermal emission from the interstellar dust in our Galaxy. The bright arc corresponds to the galactic plane.}
\label{pointing_fig}
\end{figure}

We have performed near- and mid-infrared spectroscopic observations of the zodiacal emission using AKARI/IRC in the spectroscopic mode (IRC04 in the Astronomical Observation Template). 
As part of the mission program ``SOSOS,'' we obtained data during Phase 2 (November 2006--August 2007), in which the telescope and the instruments were cooled down to 5.8 K. 
Information about the observations is listed in Table~\ref{obs_log_table}. 
Thanks to its Sun-synchronous polar orbit, AKARI has the ability to observe the entire sky in six months, although its solar elongation angle is limited to 90$\pm$1 degree. 
We observed various sky directions with 74 pointed observations in total, and Figure~\ref{pointing_fig} shows a map of those directions in ecliptic coordinates. 
They cover the whole sky except for the galactic plane, where the thermal emission from the interstellar dust in our Galaxy dominates the sky brightness in the mid-infrared~\cite{Kelsall_1998}. 

For this study, we used spectroscopic data obtained in the MIR-S channel (5--13~$\mu$m; $\lambda$/$\Delta \lambda \sim 50$), which a slit aperture with 5 arcsec width and a Si:As detector with 256 $\times$ 256 pixels are installed. This wavelength coverage is optimum for investigating the silicate features around 10~$\mu$m. 
These data were taken separately using two grisms, SG1 (5.4--8.4~$\mu$m) and SG2 (7.5--12.9~$\mu$m). 
The IRC takes four spectroscopic frames with the SG1 grism, then takes an imaging frame, and finally takes four spectroscopic frames with the SG2 grism. 
Each frame consists of a short exposure (0.5844 sec) and three long exposures (16.3632 sec). 
For each pointed observation, we analyzed the data from the long exposures in the spectroscopic frames: 12 images in total (3 long exposures $\times$ 4 spectroscopic frames) for each of SG1 and SG2.

\clearpage
\begin{longtable}[p!]{*{9}{c}}
\caption{Log for 74 pointed observations. Pointing directions were written in 3 types of celestial coordinates: equatorial coordinates ($\alpha$, $\delta$), geocentric ecliptic coordinates ($\lambda_{\oplus}$, $\beta_{\oplus}$), galactic coordinates ($l$, $b$). Solar elongations in all observations were in the range of 90$\pm$1 degree. The brightness contribution $A$ from the dust bands defined by equation (\ref{ast_contri_eq}) was also shown in the last column.}
\label{obs_log_table}\\

\hline
Observation ID & Date (UT) & $\alpha$ [deg] & $\delta$ [deg] & $\lambda_{\oplus}$ [deg] & $\beta_{\oplus}$ [deg] & $l$ [deg] & $b$ [deg] & $A$ [\%] \\ 
\hline \hline
\endfirsthead

\multicolumn{9}{l}{Continue from the previous page}\\
\hline
Observation ID & Date (UT) & $\alpha$ [deg] & $\delta$ [deg] & $\lambda_{\oplus}$ [deg] & $\beta_{\oplus}$ [deg] & $l$ [deg] & $b$ [deg] & $A$ [\%] \\ 
\hline \hline
\endhead

\hline
\multicolumn{9}{r}{Continue to the next page}\\
\endfoot

\hline
\multicolumn{9}{r}{End}\\
\endlastfoot

1500701	& 2006/12/16~18:35:01 & 174.22 & 1.41   & 174.13  & -1.00   & 	264.99  & 	58.68  & 4.90 \\
1500703	& 2006/12/20~16:55:29 & 175.02 & -7.11  & 178.27  & -8.50   & 	273.59  & 	51.65  & 1.04 \\ 
1500704	& 2006/11/22~18:56:11 & 150.22 & 3.09   & 151.20  & -8.50   & 	235.92  & 	42.71  & 1.15 \\ 
1500705	& 2007/02/04~16:40:21 & 226.81 & -7.14  & 226.40  & 10.00   &   351.68  & 	42.52  & 1.84 \\
1500706	& 2007/02/05~19:08:19 & 228.04 & -3.33  & 226.53  & 14.00   & 	356.54  & 	44.40  & 1.39 \\ 
1500707	& 2007/02/05~13:21:40 & 46.69  & 6.59   & 46.13   & -10.50  & 	172.07  & 	-43.01 & 0.82 \\
1500708	& 2007/02/05~10:01:47 & 48.07  & 2.30   & 46.27   & -15.00  & 	177.62  & 	-45.09 & 0.57 \\ 
1500709	& 2006/12/15~05:23:58 & 349.86 & 6.51   & 353.27  & 10.00   &   86.08   &  -49.66 & 1.85 \\
1500711	& 2007/02/05~14:00:26 & 239.22 & 31.05  & 227.00  & 50.00   & 	49.93   & 	49.59  & 0.79 \\ 
1500712	& 2007/02/04~19:47:57 & 239.31 & 31.03  & 227.13  & 50.00   & 	49.90   &  49.50  & 0.79 \\ 
1500713	& 2006/12/16~09:28:29 & 355.14 & -3.19  & 354.27  & -1.00   & 	84.58   &  -60.67 & 4.71 \\ 
1500715	& 2006/12/15~07:03:14 & 349.86 & 6.51   & 353.27  & 10.00   &   86.08   &  -49.66 & 1.85 \\ 
1500717	& 2006/12/17~05:50:18 & 226.05 & 60.54  & 176.40  & 70.00   & 	98.30   & 	49.76  & 0.79 \\
1500718	& 2006/12/17~07:29:34 & 226.05 & 60.54  & 176.40  & 70.00   & 	98.30   & 	49.76  & 0.79 \\
1500719  & 2006/11/10~11:15:25 & 90.00  & -66.56 & 90.00   & -90.00  & 	276.38  & 	-29.81 & 0.45 \\
1500720	& 2006/11/10~19:31:34 & 90.00  & -66.56 & 90.00   & -90.00  & 	276.38  & 	-29.81 & 0.45 \\
1500721	& 2007/02/04~21:21:44 & 248.99 & 49.66  & 227.27  & 70.00   & 	76.65   & 	41.96  & 0.77 \\
1500722	& 2007/02/03~17:13:21 & 249.05 & 49.64  & 227.40  & 70.00   & 	76.61   & 	41.92  & 0.77 \\
1500723	& 2006/12/28~13:11:21 & 230.95 & 57.48  & 188.00  & 70.00   & 	92.19   & 	49.56  & 0.81 \\
1500724	& 2006/12/28~14:50:38 & 230.95 & 57.48  & 188.00  & 70.00   & 	92.19   & 	49.56  & 0.81 \\
1500725	& 2007/02/05~14:50:05 & 59.22  & -31.05 & 47.00   & -50.00  & 	229.93  & 	-49.59 & 0.34 \\
1500726	& 2007/02/05~06:27:59 & 68.92  & -49.68 & 47.13   & -70.00  & 	256.68  & 	-42.00 & 0.33 \\
1500727	& 2006/12/19~04:59:29 & 46.71  & -60.10 & 358.00  & -70.00  & 	277.46  &  -49.79 & 0.35 \\
1500728	& 2006/12/19~06:38:46 & 46.71  & -60.10 & 358.00  & -70.00  & 	277.46  & 	-49.79 & 0.35 \\
1500729	& 2006/12/20~07:27:51 & 46.76  & -60.06 & 358.13  & -70.00  & 	277.39  & 	-49.80 & 0.35 \\
1500730	& 2006/12/20~09:07:08 & 46.76  & -60.06 & 358.13  & -70.00  & 	277.39  & 	-49.80 & 0.35 \\
1500731	& 2006/12/28~14:01:00 & 50.95  & -57.48 & 8.00    & -70.00  & 	272.19  & 	-49.56 & 0.34 \\
1500732	& 2006/12/28~15:40:16 & 50.95  & -57.48 & 8.00    & -70.00  & 	272.19  & 	-49.56 & 0.34 \\
1500748	& 2007/05/02~14:28:54 & 141.03 & 36.33  & 131.80  & 20.00   & 	187.57  & 	45.50  & 0.76 \\
1500751	& 2007/05/02~17:50:27 & 145.57 & 45.73  & 131.67  & 30.00   & 	173.52  & 	48.41  & 0.59 \\
1500759	& 2007/05/05~21:07:43 & 320.39 & -26.49	 & 314.53  & -10.50	  &   21.13   & -43.53 & 1.83 \\
1500760  & 2007/05/05~19:26:06 & 317.83 & -18.84 & 314.60  & -2.50   & 	29.97   & -39.06 & 4.71 \\
1501603	& 2007/06/17~08:56:51 & 174.34 & -3.00  & 176.00  & -5.00   &   269.39  & 54.98  & 3.09 \\
1501607  & 2007/06/19~12:17:06 & 172.52 & -12.03 &	178.00  & -14.00  &   273.55	& 46.15  & 1.52 \\
1501608  & 2007/06/19~13:56:33 & 172.52 & -12.03 & 178.00  & -14.00  &   273.55  & 46.15  & 1.52 \\
1501609  & 2007/05/30~10:09:52 & 165.66 & 22.42  & 158.00  & 15.00   &   217.73  & 64.91  & 0.66 \\
1501614  & 2007/05/31~15:55:55 & 161.83 & 9.86   & 159.50  & 2.00    &   237.54  & 56.04  & 5.09 \\
1501615	& 2007/07/13~10:01:20 & 210.04 & 19.93	 & 200.00  & 30.00   &   12.02   & 72.70  & 0.28 \\
1501617	& 2007/06/01~10:14:34 & 169.80 & 26.21	 & 160.00  & 20.00   &   210.22  & 69.40  & 0.53 \\
1501618	& 2007/06/01~13:33:25 & 169.80 & 26.21  & 160.00  & 20.00   &   210.22  & 69.40  & 0.53 \\
1501619  & 2007/08/09~17:03:30 & 239.22 & 31.05  & 227.00  & 50.00   &   49.93   & 49.59  & 0.27 \\
1501620  & 2007/08/10~11:18:01 & 239.22 & 31.05  & 227.00  & 50.00   &   49.93   & 49.59  & 0.27 \\
1501623	& 2007/06/23~11:39:56 & 2.11	   & -2.35	 & 1.00	   & -3.00	  &   98.55	& -63.17 & 4.68 \\
1501625  & 2007/07/02~13:40:07 & 193.31 & 5.70   & 190.00  & 10.50   &   304.17  & 68.57  & 0.71 \\
1501627	& 2007/07/03~12:49:22 & 189.72 & -5.27  & 191.00  & -1.00   &   297.11  & 57.46  & 4.74 \\
1501628	& 2007/07/03~14:28:50 & 189.72 & -5.27  & 191.00  & -1.00   &   297.11  & 57.46  & 4.74 \\
1501629  & 2007/06/27~11:49:36 & 7.97   & -5.82	 & 5.00	   & -8.50	  &   109.73	& -68.18 & 2.33 \\
1501631	& 2007/06/27~13:29:03 & 7.97   & -5.82  & 5.00    & -8.50   &   109.73  & -68.18 & 2.33 \\
1501633	& 2007/06/27~15:58:47 & 188.77 & 7.65   & 185.00  & 10.50   &   290.92  & 70.14  & 0.73 \\
1501635  & 2007/06/28~10:13:36 & 190.89 & 10.01  & 186.00  & 13.50   &   296.38  & 72.78  & 0.53 \\
1501639	& 2007/07/23~17:08:43 & 227.13 & 35.10  & 210.00  & 50.00   &   56.76  	& 59.76  & 0.25 \\
1501640	& 2007/07/23~23:46:38 & 227.13 & 35.10  & 210.00  & 50.00   &   56.76   & 59.76  & 0.25 \\
1501645	& 2007/06/27~18:27:33 & 8.62   & -6.08  & 5.50    & -9.00   &   111.33  & -68.57 & 2.22 \\
1501646	& 2007/06/27~20:07:00 & 8.62   & -6.08  & 5.50    & -9.00   &   111.33  & -68.57 & 2.22 \\
1501647	& 2007/06/28~17:41:00 & 11.14  & -9.35  & 6.50    & -13.00	  &   117.40  & -72.15 & 1.62 \\
1501648	& 2007/06/28~19:20:27 & 11.14  & -9.35  & 6.50    & -13.00	  &   117.40  & -72.15 & 1.62 \\
1501649	& 2007/05/13~17:11:43 & 150.43 & 29.70  & 142.00  & 16.50   &   198.71  & 52.80  & 0.77 \\
1501653	& 2007/07/02~12:42:56 & 2.41	   & 19.08  & 10.00  	& 16.50	  &   109.45	& -42.70 & 0.44 \\
1501654	& 2007/07/02~14:22:32 & 2.63  	& 18.62  & 10.00	   & 16.00	  &   109.59	& -43.19 & 0.45 \\
1501655	& 2007/07/03~18:33:12 & 3.34   & 19.47	 & 11.00 	& 16.50	  &   110.72	& -42.49 & 0.44 \\
1501660	& 2007/05/16~16:17:52 & 143.10	& -0.19	 & 145.50	& -14.00	  &   234.23	& 34.98  & 1.37 \\
1501661	& 2007/07/01~18:37:11 & 13.60	& -8.84	 & 9.00	   & -13.50  &   125.28	& -71.70 & 1.58 \\
1501662	& 2007/07/01~20:16:47 & 13.81  & -9.30  & 9.00    & -14.00  &    125.98 & -72.15 & 1.53 \\
1501663	& 2007/06/30~17:45:20 & 12.70  & -9.23  & 8.00    & -13.50	  &    122.42 & -72.10 & 1.58 \\
1501664  & 2007/06/30~12:47:06 & 12.90	& -9.69	 & 8.00	   & -14.00  &   123.08  & -72.56 & 1.53 \\
1501669	& 2007/07/18~11:43:02 & 21.47  & 13.86  & 25.00 	& 4.50	  &   135.52	& -48.17 & 1.16 \\
1501670  & 2007/07/18~13:22:31 & 21.47  & 13.86  & 25.00   & 4.50    &    135.52 & -48.17 & 1.16 \\
1501671  & 2007/07/19~00:58:59 & 22.61	& 13.77	 & 26.00	   & 4.00	  &   137.18	& -48.02 & 1.29 \\
1501672	& 2007/07/19~10:55:51 & 22.61  & 13.77  & 26.00   & 4.00    &   137.18  & -48.02 & 1.29 \\
1501673	& 2007/08/02~23:58:04 & 217.10 & -16.24 & 220.00  & -1.50   &   334.24	& 40.67  & 4.77 \\
1501675	& 2007/08/04~00:51:59 & 219.84	& -11.32 & 221.00	& 4.00	  &   340.75	& 43.48  & 1.27 \\
1501679	& 2007/08/19~12:27:54 & 240.55	& 9.97   & 236.00  & 30.00	  &    21.43	&  41.99 & 0.34 \\
1501680	& 2007/08/19~14:07:25 & 240.55	& 9.97	 & 236.00	& 30.00	  &    21.43	&  41.99 & 0.34 \\
1501685	& 2007/08/08~11:01:27 & 225.47 & -6.76  & 225.00  & 10.00   &    350.68	&  43.69 & 0.73 \\
\hline
\end{longtable}
\renewcommand{\arraystretch}{1.0}

\section{Data reduction}
\label{reduction_sec}

With a conventional reduction toolkit~\cite{Ohyama_2007}, the SG1 and SG2 spectra are not smoothly connected, owing to inconsistencies in the intensity levels in the overlapping wavelength range (i.e., 7.5--8.4~$\mu$m). 
We have examined contamination due to various instrumental artifacts, and we have identified three types of artifacts that affect the slit-spectroscopic data: (1) light scattered from the detector pixels, (2) light scattered from the edge of the detector, and (3) the ghost of a small aperture window. 
We have determined empirically normalized profiles common to all the pointing data and have calculated the absolute values from the observed images for each pointing. 
More details about these artifacts are explained in Appendix~\ref{artifacts}. 

\begin{figure}[tb]
\begin{center}
     \includegraphics[width=0.8\linewidth]{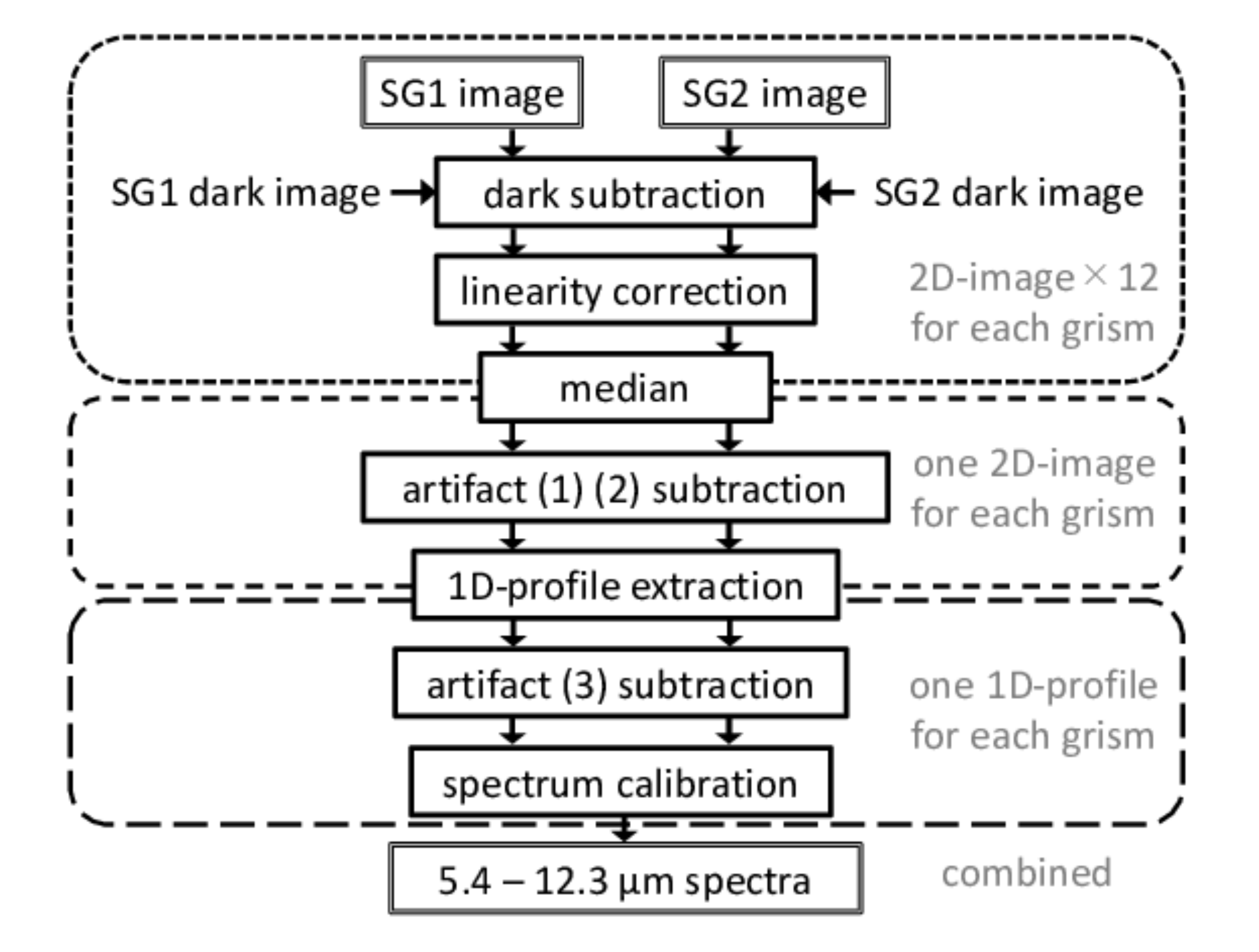}
\end{center}
\caption{The data-reduction flow we performed.}
\label{reduction_fig}
\end{figure}

The data-reduction flow, including the subtraction of these artifacts, is summarized in Figure~\ref{reduction_fig} and described below. 
For each pointing, we started from the raw data for 12 two-dimensional images taken with each grism: 3 exposures $\times$ 4 frames. 
First, we subtracted a dark image from each image. 
We obtained the dark image as the first of each set of observations. We scaled it by the median value of the dark images taken just before each pointed observation in order to correct for long-term variations of hot pixels. 
After correcting for detector linearity, we constructed a median image from these 12 images. 
This procedure removed contamination from cosmic-ray events. 
We then subtracted images representing the two scattered-light components---artifacts (1) and (2)---from the median image. 
We next extracted a one-dimensional profile from the two-dimensional image of the slit-spectroscopic region by calculating the median values of 10 pixels along the spatial direction. 
We treated the spatial variations as statistical errors. 
Assuming a Gaussian distribution, we calculated the error as the median absolute deviation scaled by a factor of 1.4826~\cite{Mosteller_1977,Tukey_1977}. 
From the extracted one-dimensional profile, we subtracted an estimated profile of artifact (3). 

The final procedure was spectrum calibration. 
We converted the pixel position and signal strength to the corresponding wavelength and intensity, respectively. 
For these conversions, we used the parameters given in Appendix 4 of Sakon~et~al.~(2007)~\cite{Sakon_2007}, which are optimized for the spectroscopic data for NGC6543, KF09T1, and HD42525, as shown in Table~\ref{cal_target_tab}. 
We derived the pixel-to-wavelength relation by using significant gas emission lines ([ArIII], [SIV], and [NeII];~\cite{Bernard-Salas_2003}) seen in the slit-spectroscopic data for NGC6543, a well-known planetary nebula, obtaining the following relations: 
\begin{eqnarray}
\lambda = 0.057 \times (Y-12) + 0.016   \mbox{~~~for SG1,}\\
\lambda = 0.099 \times (Y-31) + 0.141   \mbox{~~~for SG2,} 
\label{wlcal_eq}
\end{eqnarray}
where $\lambda$ is the corresponding wavelength in units of~$\mu$m, and $Y$ is the pixel position along the direction of spectral dispersion ($1 \leq Y \leq 256$). 
To calibrate the intensity at each wavelength, we derived a response curve by referring to the spectroscopic data for two standard stars, KF09T1 and HD42525, and using the model spectra provided by M. Cohen and coworkers (e.g., \cite{Cohen_1995, Cohen_1996, Cohen_1999, Cohen_2003}). 
Since the AKARI spectra of these stars were taken in the main field of view without a slit mask, we took account of the uncertainties in the reference-source positions ($< \pm$0.5 pixel)---which result in wavelength shifts of the reference spectra ($< \pm0.0285$~$\mu$m and $\pm0.0496$~$\mu$m for SG1 and SG2, respectively)---as statistical errors in the response curve. 

\begin{table}[tb]
  \centering
  \caption{List of the data used for spectrum calibration.}
  \begin{tabular}{cccccc} \hline
    Object   &  Object type  &  Observation ID  &   file   &  spectroscopic mode  \\ \hline
    NGC6543  &  planetary nebula  &  5020048  &  F33860-M.fits  &  slit spectroscopy \\ 
    KF09T1  &  K0III-type star  &  5020032  &  F13450-M.fits  &  slit-less spectroscopy \\ 
    HD42525  &  A0V-type star  &  5020023  &  F23384-M.fits  &  slit-less spectroscopy \\ \hline
  \end{tabular}
  \label{cal_target_tab}
\end{table}

\begin{figure}[htb]
\begin{center}
     \includegraphics[width=0.88\linewidth]{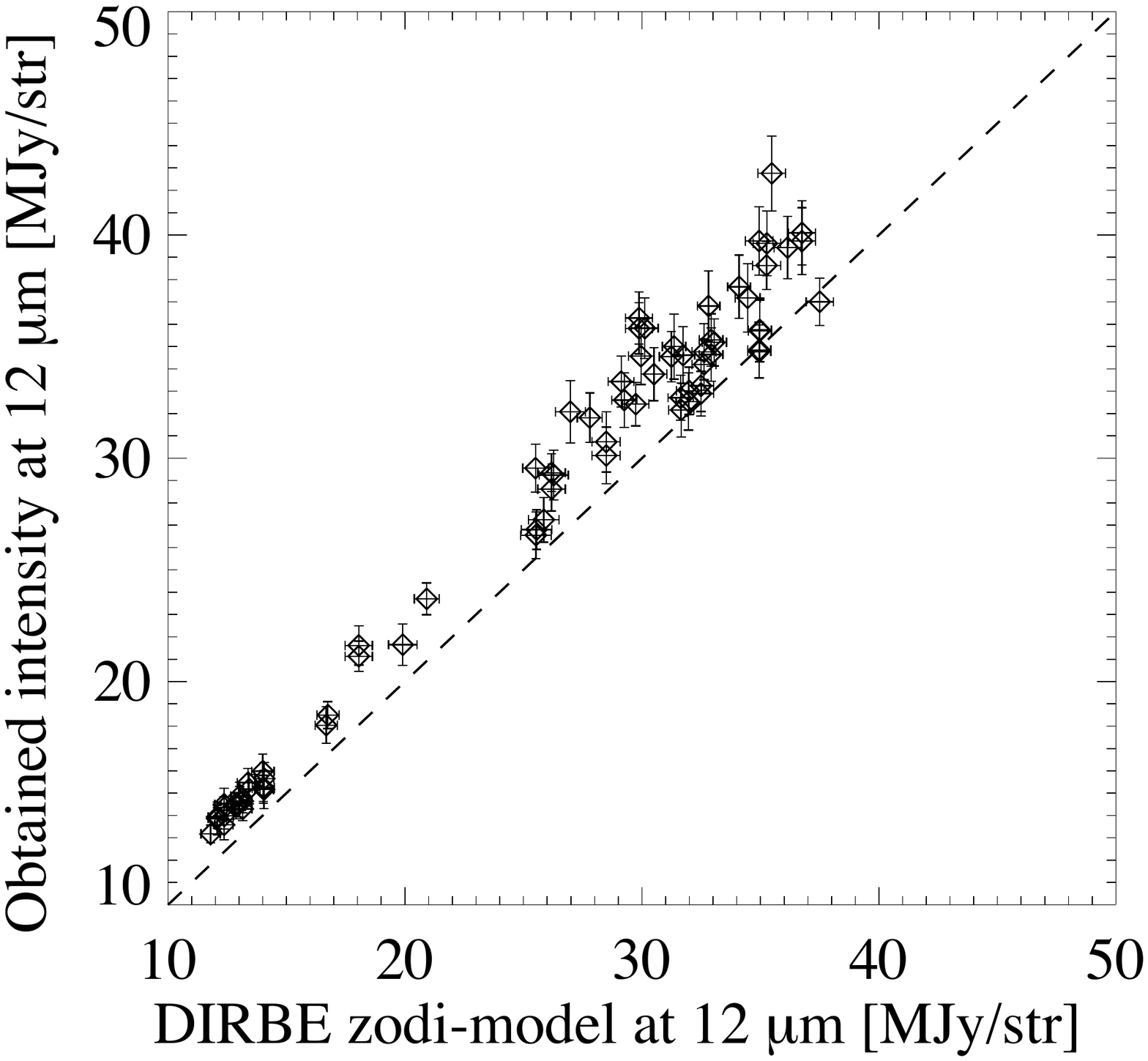}
\end{center}
\caption{Correlation of the 12~$\mu$m intensity between the spectra obtained in this work and the DIRBE zodi-model~\cite{Kelsall_1998}.}
\label{DIRBE-AKARI_fig}
\end{figure}

In summary, we were able to subtract the three types of artifact components and derive reasonable spectra, smoothly connecting between the data from SG1 and SG2 for all pointing directions. 
The spectra include two types of statistical errors: those due to spatial dispersion during the extraction of the one-dimensional profile and those due to uncertainties in the response curves we used. 
The intensity levels are consistent within 10\% accuracy with those calculated from the zodiacal-emission model obtained from DIRBE imaging observations (hereafter, the ``DIRBE zodi-model:'' \cite{Kelsall_1998}). 
Figure~\ref{DIRBE-AKARI_fig} shows the correlation of the intensities at 12~$\mu$m between the AKARI observations used in this work and the DIRBE zodi-model. 
The AKARI intensities are about 10\% higher than the model expectations uniformly at all sky directions. 
One of reasons for the 10\% inconsistency may be an inappropriate color-correction factor used for the model construction from the DIRBE observational data. 
Kelsall~et~al.~(1998)~\cite{Kelsall_1998} assumed a color-correction factor appropriate for a single-temperature blackbody over the 8.8--15.2~$\mu$m band. 
However, actual zodiacal spectra seem to have deviations from a blackbody due to excess emission in the range 9--12~$\mu$m, which is found in this AKARI observation, and a deficit in emission at the side longer than 12~$\mu$m on the contrary, which is expected from the decline of absorption coefficients (i.e., emission efficiencies) of small silicate grains~\cite{Koike_2003, Chihara_2002}. 
Since the DIRBE response in this wavelength band is more sensitive at the longer side, the 12~$\mu$m-intensity may be resulted in a lower one than the actual. 
Another possibility causing the 10\% inconsistency is the contribution of an isotropic component with an brightness uniform all over the sky, which comes also from an external to the solar system. 
The DIRBE zodi-model ignores such isotropic component and is fitted so that it can reproduce the amplitude and phase of the temporal variation from the average for each line of sight (equation (10) of \cite{Kelsall_1998}). 

\section{Results}
\label{result_sec}

\subsection{Detection of emission features}

We found the emission features around 10~$\mu$m to be present in the spectra at all pointing directions. 
Hereafter, we use the spectral data obtained from the SG2 (7.5--12.9~$\mu$m) to investigate the detailed feature shapes, while the SG1 data have been used to confirm the subtraction of the artifact components. 
The SG2 spectral data show the first reliable feature shapes around 9--10~$\mu$m. 
ISOCAM spectra contained the discontinuity at 9.5 $\mu$m across the boundaries of the spectra obtained from two different circular-variable filters~(see Figure 4 of \cite{Reach_2003}), and this resulted in the uncertainty of the feature shapes especially in the shorter-wavelength side. 
On the other hand, the SG2 in the AKARI/IRC covers the entire wavelength region (8--12~$\mu$m) of the emission features and we can detect the feature shapes by using spectra obtained only from the single spectroscopic element. 

In order to extract the excess emission first of all, we divided observed spectra by the continuum spectrum in each direction. 
We calculated the continuum spectra from the DIRBE zodi-model mentioned above. 

In the mid-infrared wavelength region, the intensity of the zodiacal emission can be written as follows: 
\begin{equation}
Z_{\lambda}(\lambda_{\oplus},\beta_{\oplus},t) = \sum_{c=1}^{3} \int n_{c}(\lambda_{\oplus},\beta_{\oplus},s,t) E_{c,\lambda} B_{\lambda}(T(\lambda_{\oplus},\beta_{\oplus},s,t)) ds , 
\label{ZE_eq}
\end{equation}
where the subscript $c$ corresponds to the type of IPD component defined in the DIRBE zodi-model~\cite{Kelsall_1998}: a smooth cloud ($c=1$), dust bands ($c=2$), and a circumsolar ring with a trailing blob ($c=3$), which have distinct spatial distributions. 
The quantity $n_{c}$ is the spatial density of the geometrical cross-section for each IPD component. 
Positions in the Solar System are specified by the geocentric ecliptic coordinates of the pointing direction, $(\lambda_{\oplus},\beta_{\oplus})$, the distance $s$ along the line of sight, and the observation date $t$. 
The quantity $E_{c,\lambda}$ is the emissivity-modification factor, which is a function of wavelength and corresponds to deviations from the blackbody function ($B_{\lambda}(T)$). 
Here $T$ is the typical temperature of IPD grains at each position. 
Kelsall~et~al.~(1998)~\cite{Kelsall_1998} determined $n_{c}$ and the spatial distribution of $T$ by fitting to the DIRBE imaging observational data using the color-correction factors. 

We obtained the continuum spectra from these results. 
The modeled absolute intensities of the zodiacal emission, however, include about 10\% uncertainties due to the inconsistency described in Figure~\ref{DIRBE-AKARI_fig}. 
We therefore scaled the calculated continuum intensity to match the observed intensity at 12~$\mu$m. 
Denoting the scaling ratio by $\alpha$, we express the continuum spectrum $C_{\lambda}$ as
\begin{equation}
C_{\lambda}(\lambda_{\oplus},\beta_{\oplus},t) = \alpha \times \sum_{c=1}^{3} \int n_{c}(\lambda_{\oplus},\beta_{\oplus},s,t) B_{\lambda}(T(\lambda_{\oplus},\beta_{\oplus},s,t)) ds . 
\label{continuum_eq}
\end{equation}
The value of $\alpha$ was typically in the range from 1.0 to 1.1. 

We divided the observed zodiacal emission spectra $Z_{\lambda}$ by the continuum $C_{\lambda}$ for each of the 74 observations (see Appendix \ref{allemissivity}). 
We discuss below the wavelength dependence of these observed/continuum ratios, i.e., the shapes of the spectral features. 
Although the shapes exhibit diversity, the excess emission---and even some sharp emission peaks---can be seen clearly beyond the error range in all of the spectra in the 8--12~$\mu$m region. 
Such detailed shapes have been found here for the first time thanks to the high sensitivity of the IRC and the accurate subtraction of instrumental artifacts.

\subsection{Absorption coefficients of candidate minerals}

We next link the shapes of the emission features to the grain properties of the IPD included in the line of sight, from a mineralogical point of view. 

The observed/continuum spectra we obtained can be written as follows, using equations (\ref{ZE_eq}) and (\ref{continuum_eq}): 
\begin{equation}
\frac{Z_{\lambda}}{C_{\lambda}} 
= \frac{1}{\alpha} \times \frac{ \sum_{c=1}^{3} \int n_{c} E_{c,\lambda} B_{\lambda}(T) ds } { \sum_{c=1}^{3} \int n_{c} B_{\lambda}(T) ds }~.
\label{emissivity_eq}
\end{equation}
This means the observed/continuum ratios correspond to the emissivity-modification factor $E_{c,\lambda}$, averaged over all the IPD grains included in the line of sight, which have various compositions and grain sizes. 
The value of $E_{c,\lambda}$ for each grain can be replaced by the grain absorption coefficient $Q_\mathrm{abs}$. 
The observed/continuum ratios are thus determined by the $Q_\mathrm{abs}$ values of all the IPD grains included in the line of sight. 
Since $Q_\mathrm{abs}$ for a grain can be expressed approximately as a superposition of the $Q_\mathrm{abs}$ values for the constituent materials in the grain~\cite{Bohren_1977, Bohren_1983}, comparison of the observed/continuum ratios to the $Q_\mathrm{abs}$ values for candidate minerals helps us to estimate the fractional content of each type of mineral in the grain. 

\begin{figure}[p!]
\begin{center}
     \includegraphics[width=0.6\linewidth]{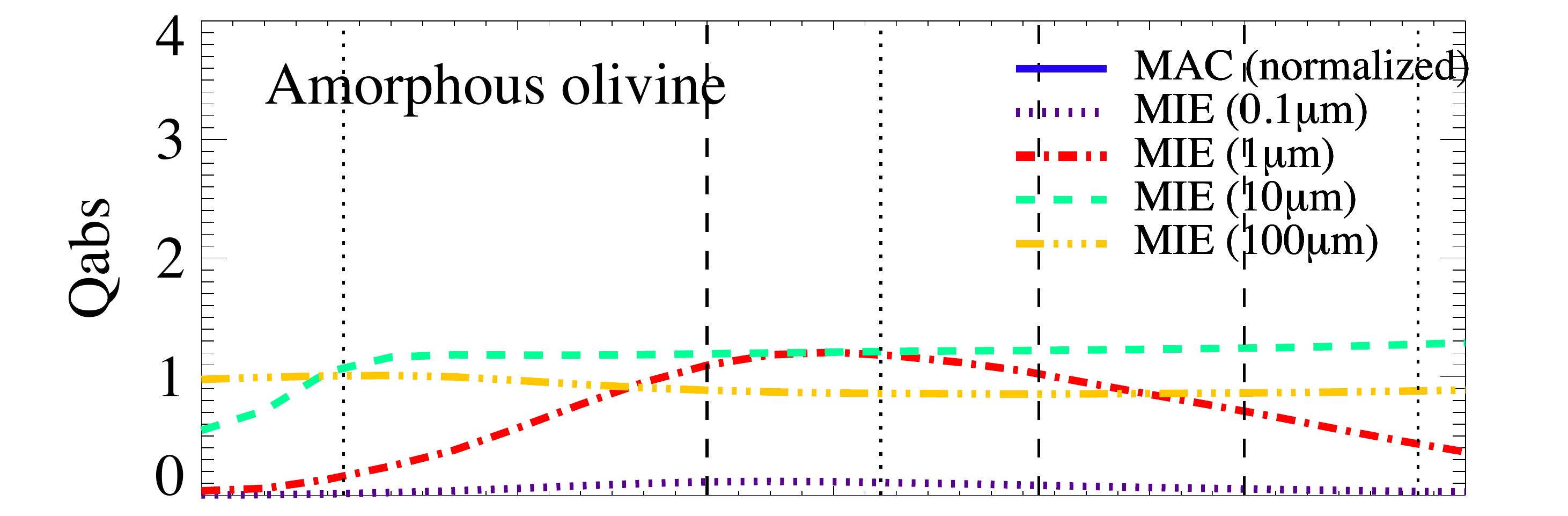}
     \includegraphics[width=0.6\linewidth]{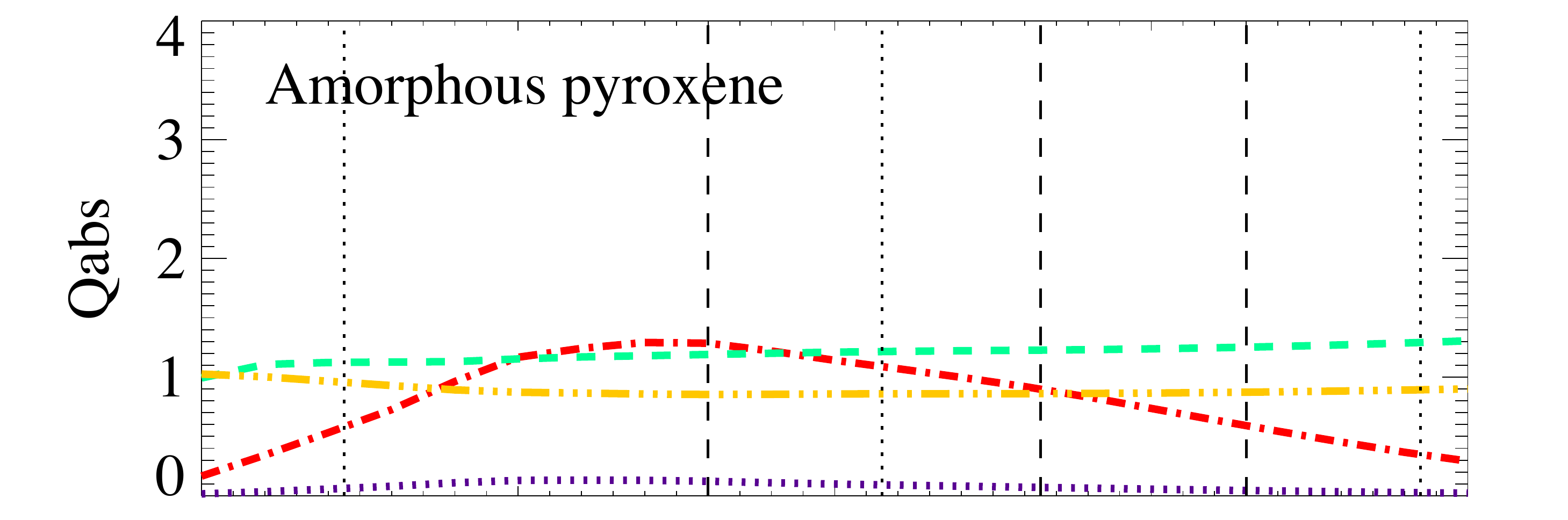}
     \includegraphics[width=0.6\linewidth]{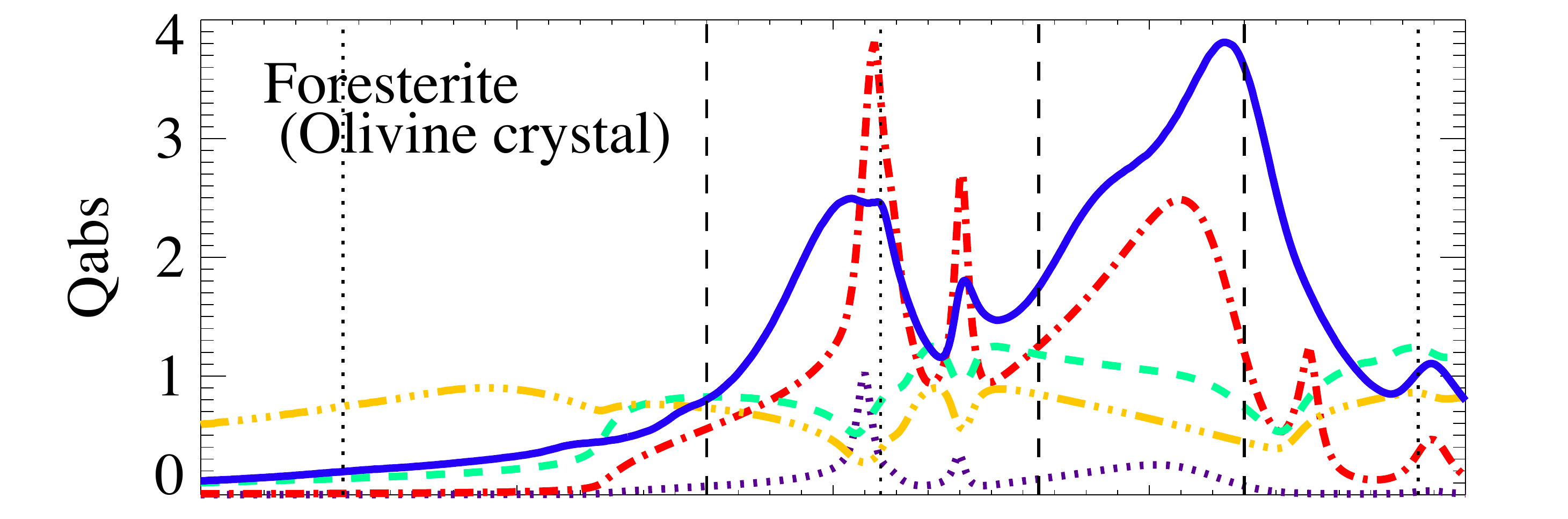}
     \includegraphics[width=0.6\linewidth]{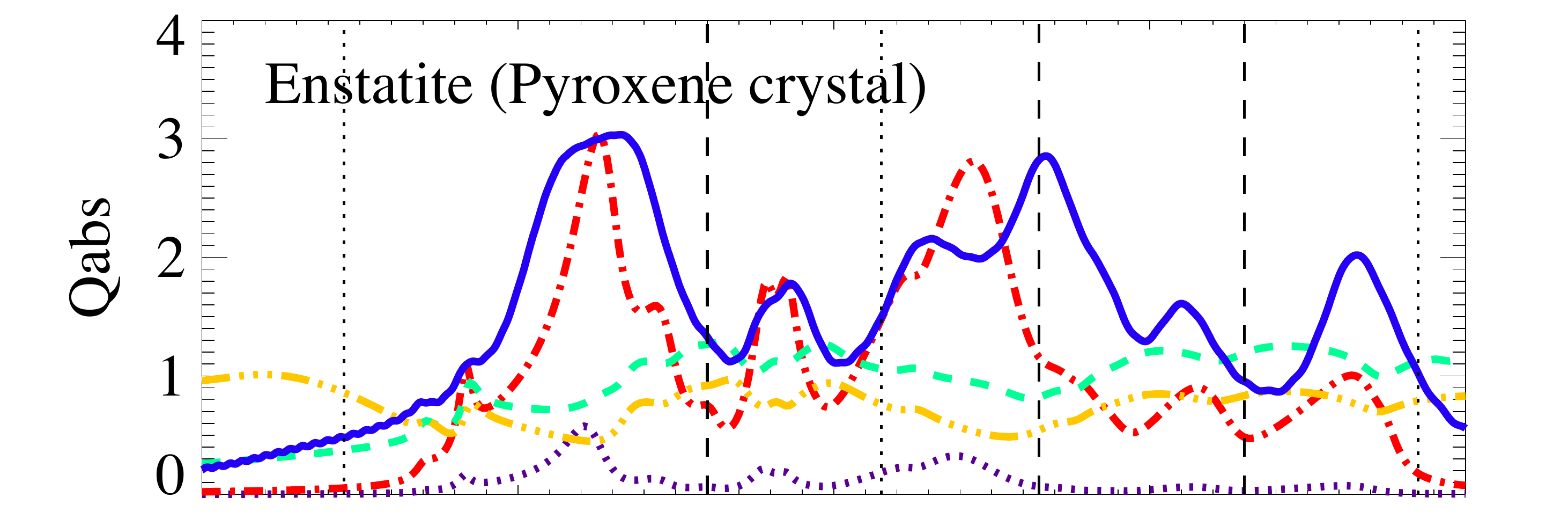}
     \includegraphics[width=0.6\linewidth]{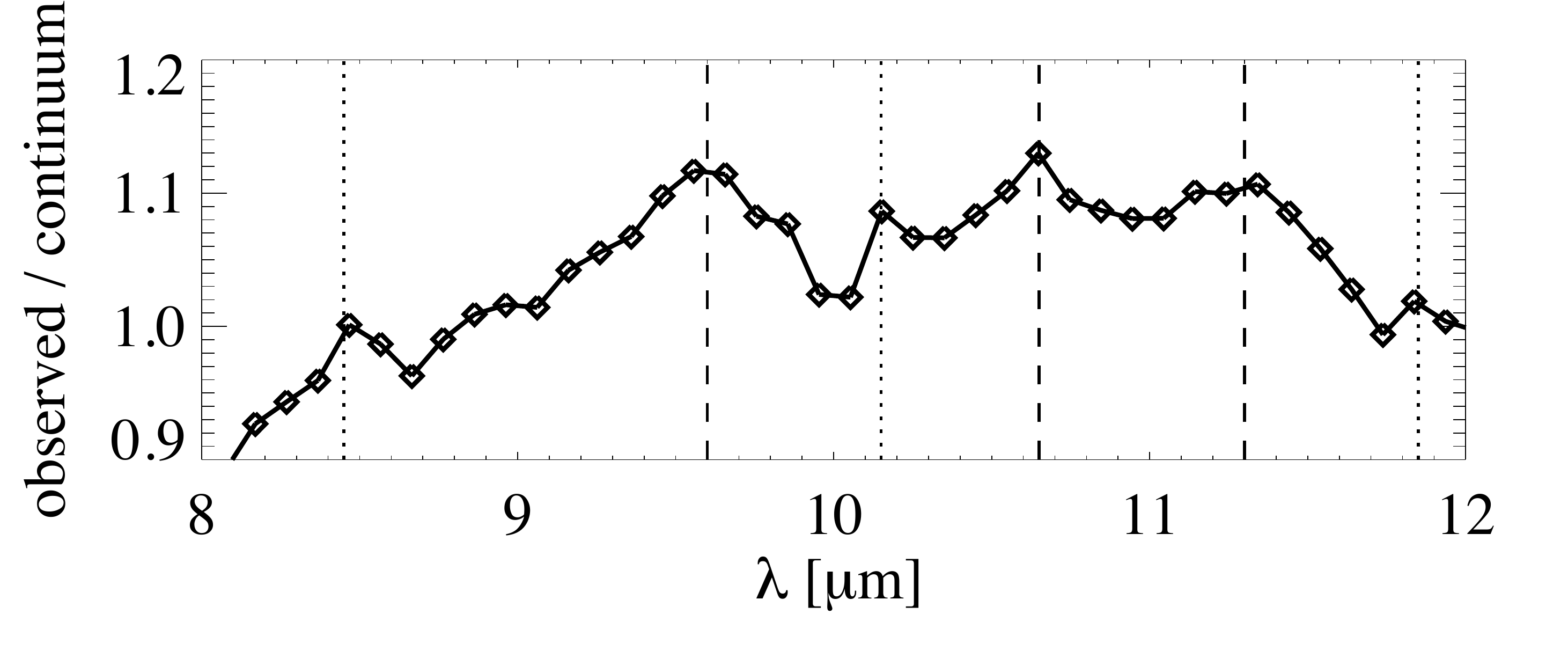}
\end{center}
\caption{Absorption coefficients of candidate minerals. Top to bottom: amorphous with olivine composition, amorphous with pyroxene composition, Mg-end forsterite (Mg$_{2}$SiO$_{4}$, one of the olivine crystals), and Mg-end ortho-enstatite (MgSiO$_{3}$, one of the pyroxene crystals). For comparison, the bottom-most panel shows the observed/continuum spectrum of the zodiacal emission averaged over all directions. We assumed a solid solution, with Mg/(Mg+Fe)=0.5 for the two amorphous minerals. We calculated absorption coefficients from their optical constants or dielectric functions on the basis of Mie theory, assuming grain sizes of 0.1, 1, 10, and 100~$\mu$m. We obtained the optical constants of the two amorphous minerals from Dorschner~et~al.~(1995)~\cite{Dorschner_1995}, and the dielectric functions of forsterite and enstatite from Sogawa~et~al.~(2006)~\cite{Sogawa_2006} and from private communications with H. Chihara. The solid blue lines in the middle two panels represent the mass absorption coefficients (MACs) measured in the laboratory~\cite{Koike_2003, Chihara_2002}, with the peaks scaled to the maximum value of $Q_\mathrm{abs}$ calculated from Mie theory, assuming 1~$\mu$m-sized grains.}
\label{Qabs_fig}
\end{figure}

In this paper, we consider four types of silicates as candidates: amorphous with olivine composition, amorphous with pyroxene composition, Mg-end forsterite (Mg$_{2}$SiO$_{4}$, one of the olivine crystals) and Mg-end ortho-enstatite (MgSiO$_{3}$, one of the pyroxene crystals). 
We assume solid solutions of the two types of amorphous minerals, with Mg/(Mg+Fe)=0.5. 
These candidate minerals are among the main constituents of IPD samples collected in the stratosphere or on the ground~\cite{Messenger_2013, Keller_2005}. 
In addition, they have vibration modes in the wavelength region we have considered in this work. 
We calculated their $Q_\mathrm{abs}$ spectra from the optical constants~\cite{Dorschner_1995} or dielectric functions (Sogawa et al. 2006~\cite{Sogawa_2006} and private communications with H. Chihara) on the basis of Mie theory~\cite{Bohren_1983, Kerker_1969}, assuming spherical grains with a size variation from 0.1~$\mu$m to 100~$\mu$m. 
The top four panels in Figure~\ref{Qabs_fig} show the results. 
We also plotted the mass absorption coefficients (MACs) for two types of crystals measured in the laboratory~\cite{Koike_2003, Chihara_2002}. 
The MACs were measured on crystalline dust samples with a size range 0.1--10~$\mu$m. 
These samples are assumed to have the shape probability distribution of a continuous distribution of ellipsoids (CDE), which includes ellipsoidal grains with equal probabilities of all possible aspect ratios and orientations of the crystal axes. 

Large grains ($\geq$ 10~$\mu$m) contribute mainly to the absolute intensity level of the zodiacal emission and provide a blackbody emission baseline~\cite{Reach_2003}, because their $Q_\mathrm{abs}$ values are relatively constant---around unity---compared to those for smaller grains, as you can see in Figure~\ref{Qabs_fig}. 
In addition to this baseline, small grains ($\leq$ 1~$\mu$m) can produce excess emission at specific wavelengths depending on their vibration modes. 
The spectral shape does not depend on the grain size regarding such small grains. 
Therefore, our study of the shapes of the excess emission features can reveal the properties of small IPD grains, which have not altered well by space weathering (see section \ref{intro}). 

\subsection{Interpretation of the average feature shapes}
\label{averaged_feature_sec}

In order to understand the typical properties of the small IPD grains over the entire sky, we first averaged the observed/continuum spectra over all 74 observations and compared the results with the $Q_\mathrm{abs}$ values of candidate minerals. 
In the average observed/continuum spectrum shown in the bottom panel of Figure~\ref{Qabs_fig}, we found three main peaks---around 9.60, 10.65, and 11.30~$\mu$m---with sub-peaks around 8.45, 10.15, and 11.85~$\mu$m. 
Such sharp peaks can be caused only by small grains ($\leq$ 1~$\mu$m) of crystalline silicates. 
We can therefore say that the IPD typically includes some fraction of such small grains of crystalline silicates. 
In particular, the main peaks around 9.60 and 10.65~$\mu$m can be caused by enstatite, while the sub-peaks around 10.15 and 11.85~$\mu$m seem to be mainly contributed by forsterite. 
Around 11.30~$\mu$m, both forsterite and enstatite show peaks. 
The sub-peak at 8.45~$\mu$m may originate from materials other than those shown in Figure~\ref{Qabs_fig}. 
Since the peaks originating from enstatite are more prominent than those from forsterite and other materials, the IPD appears to typically include plenty of enstatite. 

We note that the wavelength position of the enstatite peak around 9.2~$\mu$m needs to be shifted to around 9.6~$\mu$m, in order to reproduce the obtained observed/continuum spectrum. 
Several factors can cause wavelength shifts of this peak. 

\begin{figure}[!b]
\begin{center}
     \includegraphics[width=1\linewidth]{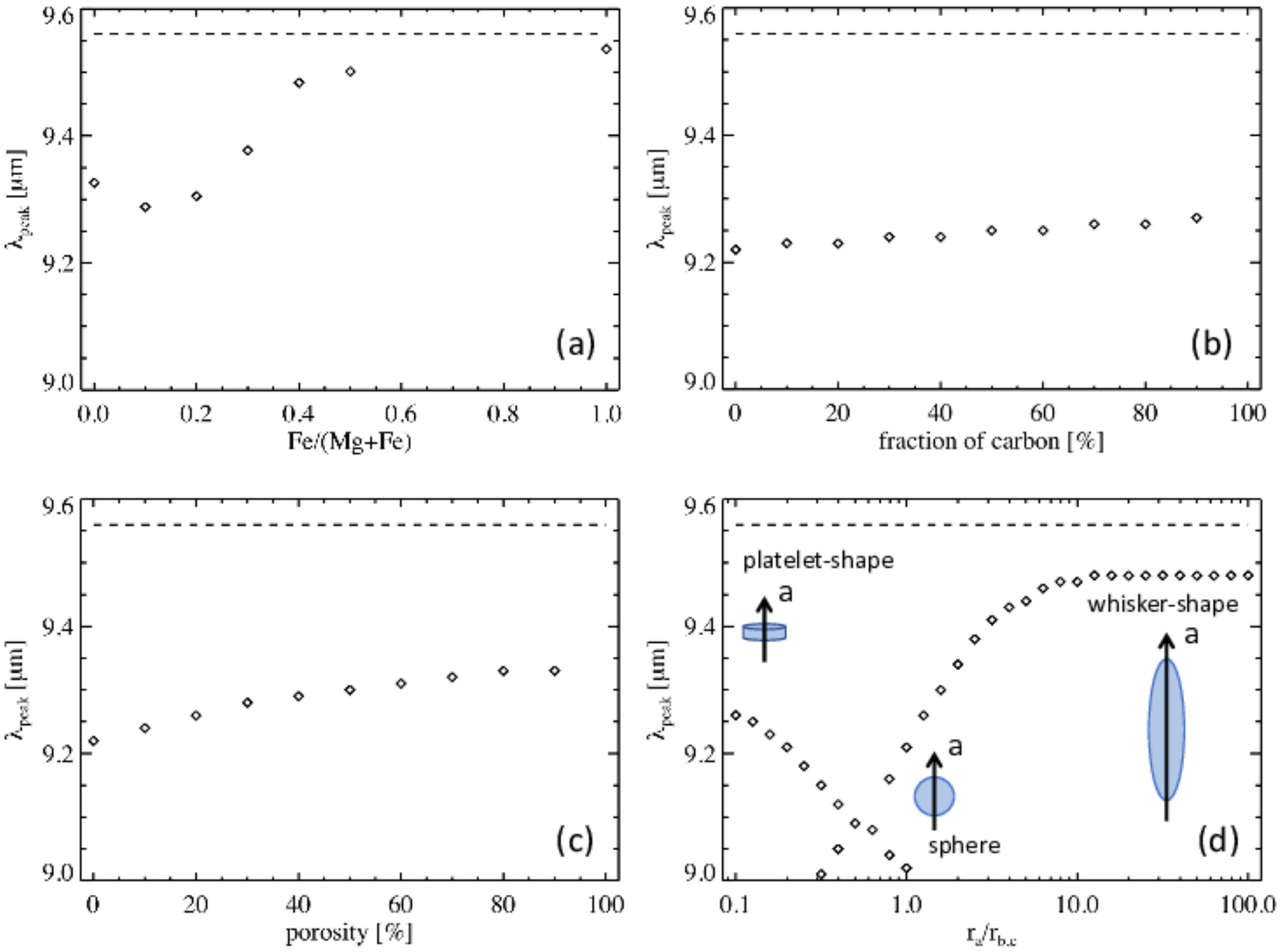}
\end{center}
\caption{The wavelength shift of a peak around 9.2~$\mu$m due to (a) replacing Mg-ions with Fe-ions, (b) coating by carbon, (c) increasing porosity, and (d) elongation along the crystal a-axis. Here, $\lambda_{peak}$ means the peak wavelength, and dotted lines at $\lambda_{peak}$ = 9.56~$\mu$m indicate the peak position in the zodiacal emission spectra. The values in panel (a) were extracted from the MACs for solid solutions, which were measured for samples in the size range 0.1--10~$\mu$m. In the calculations for panels (b) and (c), we assumed the grain size to be 1~$\mu$m. For panel (d), we fixed the grain lengths along the b- and c-axes ($r_{b,c}$) at 1~$\mu$m, and we changed the length along the a-axis ($r_{a}$) from 0.1~$\mu$m to 100~$\mu$m.}
\label{ws_fig}
\end{figure}

For example, replacing Mg-ions with Fe-ions is one possible reason for the shift of a peak to a longer wavelength, although the trend is slightly different for different peaks~\cite{Koike_2003, Chihara_2002}. 
Using the experimental results for the MACs from Chihara~et~al.~(2002)~\cite{Chihara_2002}, we plot the wavelengths of a peak around 9.2~$\mu$m in various solid solutions of enstatite as panel (a) in Figure~\ref{ws_fig}. 
The wavelength position shifts to the longer side as the Fe/(Mg+Fe) ratio increases. 
Fe-end ferrosillite (FeSiO$_{3}$) shows a peak at 9.54~$\mu$m, which is close to the observed peak position. 

\begin{figure}[tb]
\begin{center}
     \includegraphics[width=0.85\linewidth]{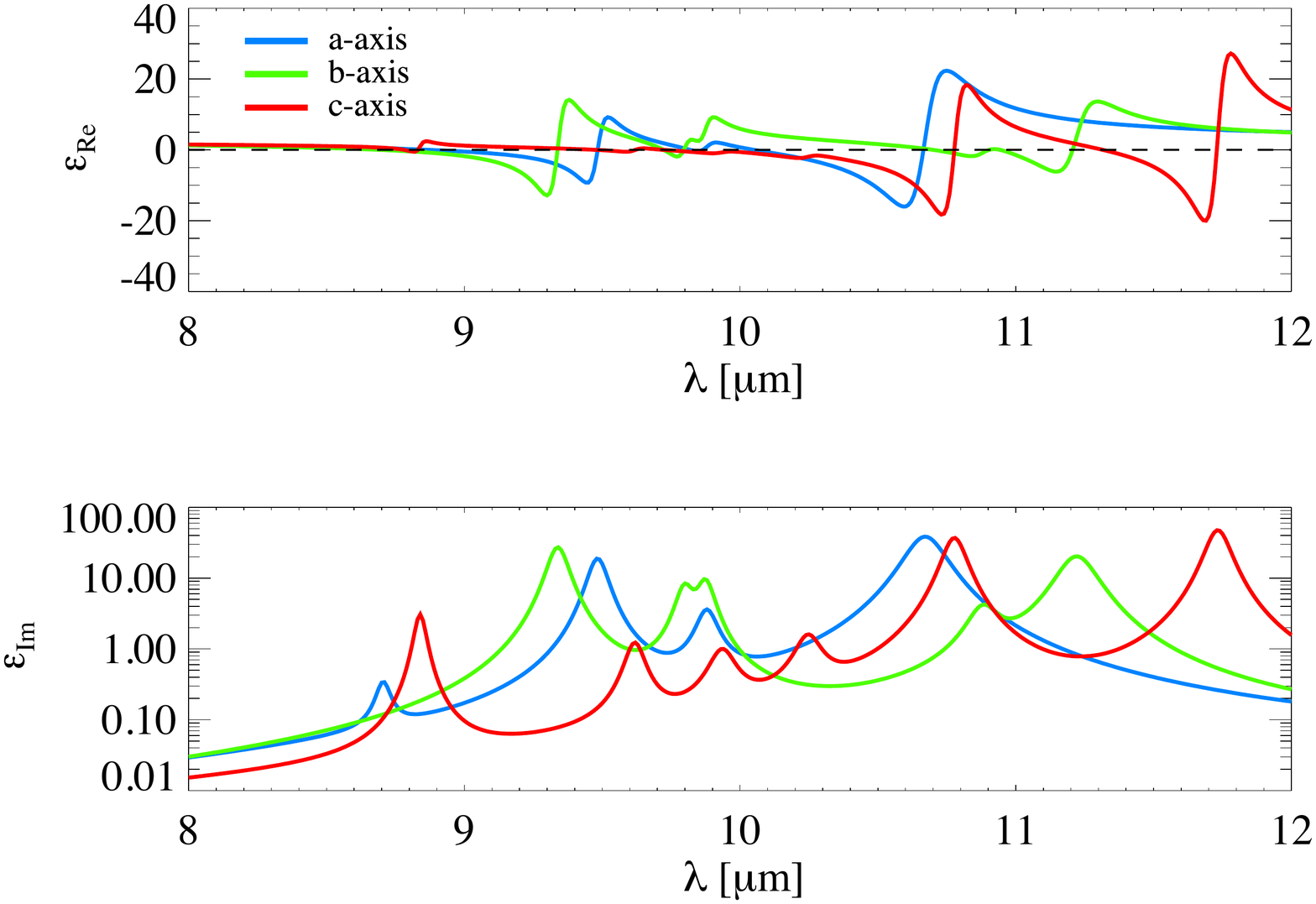}
\end{center}
\caption{Dielectric function of orthorhombic-enstatite. The top and bottom panels show the real  and imaginary parts, respectively. The c-axis is parallel to the direction of the tetrahedral chains, and the a-axis corresponds to the stacking direction of the layers of tetrahedral chains.}
\label{epsilon_fig}
\end{figure}

Grains coated by organic materials also can show peaks at shifted wavelength positions~\cite{Kimura_2013}. 
This effect can occur at peaks for which the real part of the dielectric function shows a negative valley~\cite{Bohren_1983}, as does the enstatite peak around 9.2~$\mu$m (see Figure~\ref{epsilon_fig}). 
We assumed grains with a carbon-matrix and spherical enstatite-inclusions and calculated the dielectric function of the mixture following the Maxwell-Garnett law (e.g., \cite{Bohren_1983, Genzel_1973, Barker_1973, Bohren_1977}). 
We then derived the $Q_\mathrm{abs}$ spectrum from such composite dielectric functions using Mie theory~\cite{Bohren_1983, Kerker_1969}. 
Changing the fraction of inclusions, we investigated the shift of the peak wavelengths. 
The result is shown in panel (b) of Figure~\ref{ws_fig}. 
The carbon mantle certainly shifts the peak wavelength up to 9.28~$\mu$m, but it is not sufficient to explain the observed peak wavelength. 

Similarly, the wavelength position of the peak depends on the porosity of grains. 
We assumed a porous grain consisting of an enstatite matrix with spherical vacuum inclusions. 
We calculated the $Q_\mathrm{abs}$ spectrum in the same way as for grains coated by carbon. 
The peak is shifted to longer wavelengths by increasing the porosity, as in panel (c) of Figure~\ref{ws_fig}. 
An extremely porous grain exhibits a peak at 9.33~$\mu$m, but this is again insufficient to explain the observations. 

Another possibility is the effect of crystal morphology. 
Ortho-enstatite is an orthorhombic crystal, and each crystal axis (a, b, c) has an individual dielectric function (see Figure~\ref{epsilon_fig}). 
If we consider an ellipsoidal grain, the $Q_\mathrm{abs}$ spectrum is determined by the dielectric functions of each crystal axis and the aspect ratio in the directions of the crystal axes, assuming the surface mode in the Lorentz model~\cite{Bohren_1983, Takigawa_2012}. 
Therefore, the wavelength positions of some peaks can be shifted depending on the aspect ratio~\cite{Takigawa_2012}. 
This effect also can occur at peaks for which the real part of the dielectric function shows a negative valley~\cite{Bohren_1983}. 
An example of enstatite grains flattened or elongated along the a-axis (the stacking direction of the layers of tetrahedral chains) is shown in panel (d) of Figure~\ref{ws_fig}. 
We calculated the $Q_\mathrm{abs}$ spectra of such ellipsoidal grains on the basis of the Lorentz model. 
We fixed the grain lengths along the b- and c-axes at 1~$\mu$m and changed the length along the a-axis from 0.1~$\mu$m to 100~$\mu$m. 
As shown in Figure~\ref{ws_fig}, there are two separate peaks in the shorter half of 9~$\mu$m band. 
One curve descending toward the right shows the wavelength position of a peak contributed by the b-axis, and another curve, which is increasing and asymptotic, shows that of a peak contributed by the a-axis. 
The wavelength position of the latter peak dramatically increases as a grain is elongated along the a-axis and finally, asymptotically approaches 9.48~$\mu$m. 

No single effect alone is sufficient to shift the peak wavelength to reproduce the observed peak position, but the wavelength shift may be attributed to a combination of several effects. 
These effects are reasonable, because they are consistent with the results of laboratory measurements on collected samples of the IPD. 
A lot of the collected IPD samples were found to be porous and/or carbonaceous~\cite{Messenger_2013, Bradley_1999}, and many whisker- or platelet-shaped crystals have actually been found in those samples~\cite{Bradley_1983}. 

In addition, by comparing the observed/continuum spectrum and the $Q_\mathrm{abs}$ spectra of enstatite, we found that the enstatite peak at 11.6~$\mu$m is suppressed. 
According to Figure~\ref{epsilon_fig}, the dielectric functions of the a- and b-axes are flat at wavelengths longer than 11.5~$\mu$m, although the c-axis does have a significant peak in this wavelength range~\cite{Demichelis_2012}. 
The observed suppression implies that the grain length along the c-axis (the direction of the tetrahedral chains) may be reduced, corresponding either to platelet-shaped enstatite that is symmetric around the c-axis or to whisker-shaped enstatite that is elongated along the a- or b-axes. 

\subsection{Variations among different sky directions}
\label{direction_compare_sec}

Since the observed/continuum spectra in Appendix \ref{allemissivity} show a variety of feature shapes, we also compared the feature shapes among the various sky directions. 
We first considered the effect on the feature shapes due to dust bands. 
The brightness contribution from dust bands in each direction was determined from the spatial distribution of the IPD given by the DIRBE zodi-model. 
This model includes three dust bands---around $\pm 1^{\circ}.4$, $\pm 10^{\circ}$, and $\pm 15^{\circ}$---as one of the IPD components~\cite{Kelsall_1998}. 
For each observed line of sight, we calculated the continuum intensity at 12~$\mu$m originating from the dust band components, defining $A$ as the ratio to the continuum intensity at 12~$\mu$m originating from all the components. 
This is given by 
\begin{equation}
A = \frac{\int n_{2} B_{12\mathrm{\mu m}}(T) ds}{\sum_{c=1}^{3} \int n_{c} B_{12\mathrm{\mu m}}(T) ds} \times 100 ~~~[\%]~, 
\label{ast_contri_eq}
\end{equation}
where the definitions of all the variables are the same as in equation (\ref{ZE_eq}), and the component represented by $c = 2$ corresponds to the dust bands. 

\begin{figure}[htbp]
\begin{center}
     \includegraphics[width=\linewidth]{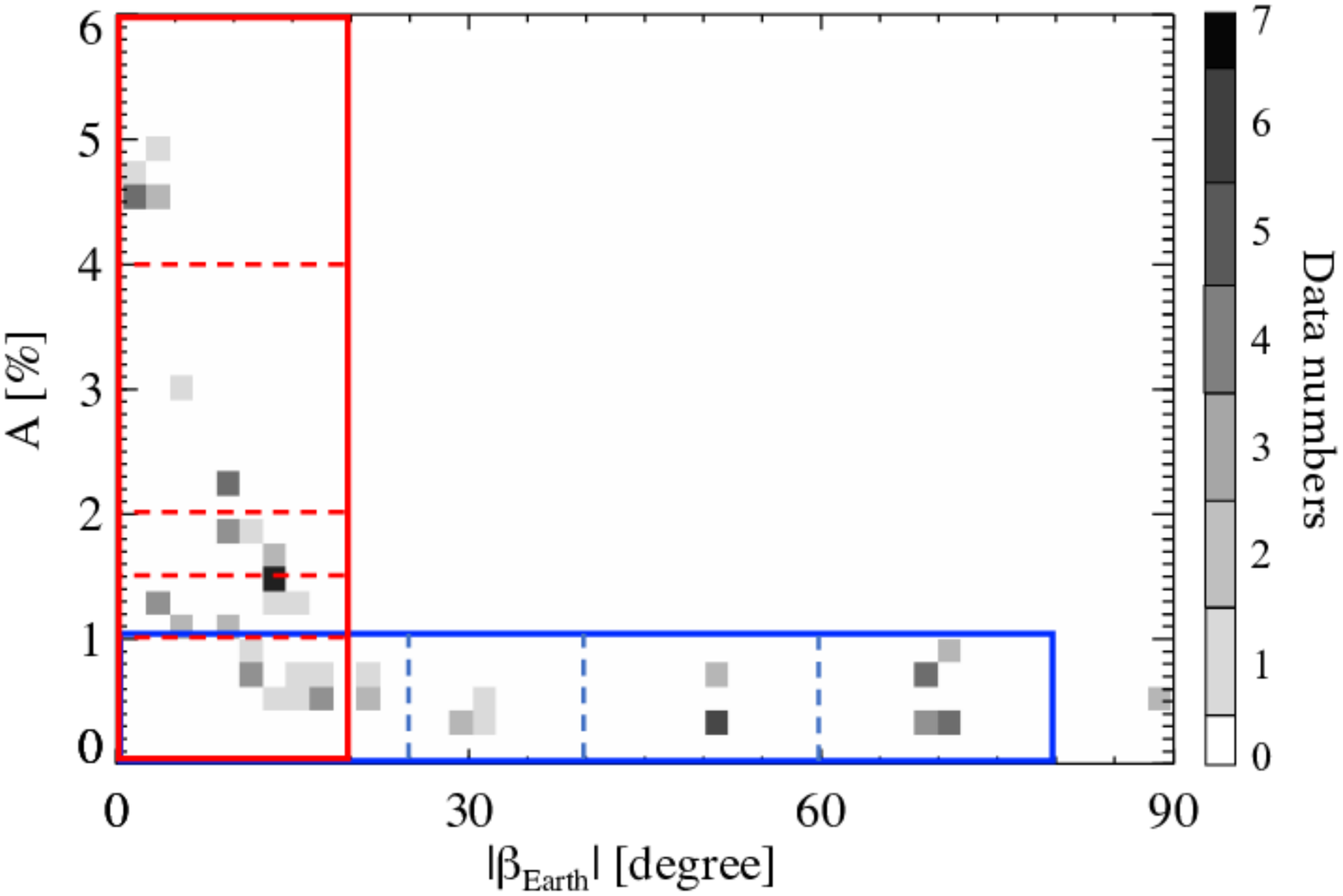}
\end{center}
\caption{Two-dimensional histogram showing the brightness contribution $A$ from the dust bands vs. the absolute values $|\beta_{\oplus}|$ of the ecliptic latitudes in the various pointing directions. We compared the feature shapes among the values of $A$ in the datasets with $|\beta_{\oplus}| < 20^{\circ}$, enclosed by the red line in the figure, and investigated the dependence of the feature shapes on $|\beta_{\oplus}|$ in the datasets with $A < 1 \%$, enclosed by the blue line. Dotted lines indicate the borders of the bins used for the averaging. We did not use the data at $|\beta_{\oplus}| = 90^{\circ}$ because the number of data points was too small.}
\label{2d_hist_fig}
\end{figure}

\begin{figure}[htbp]
\begin{center}
     \includegraphics[width=\linewidth]{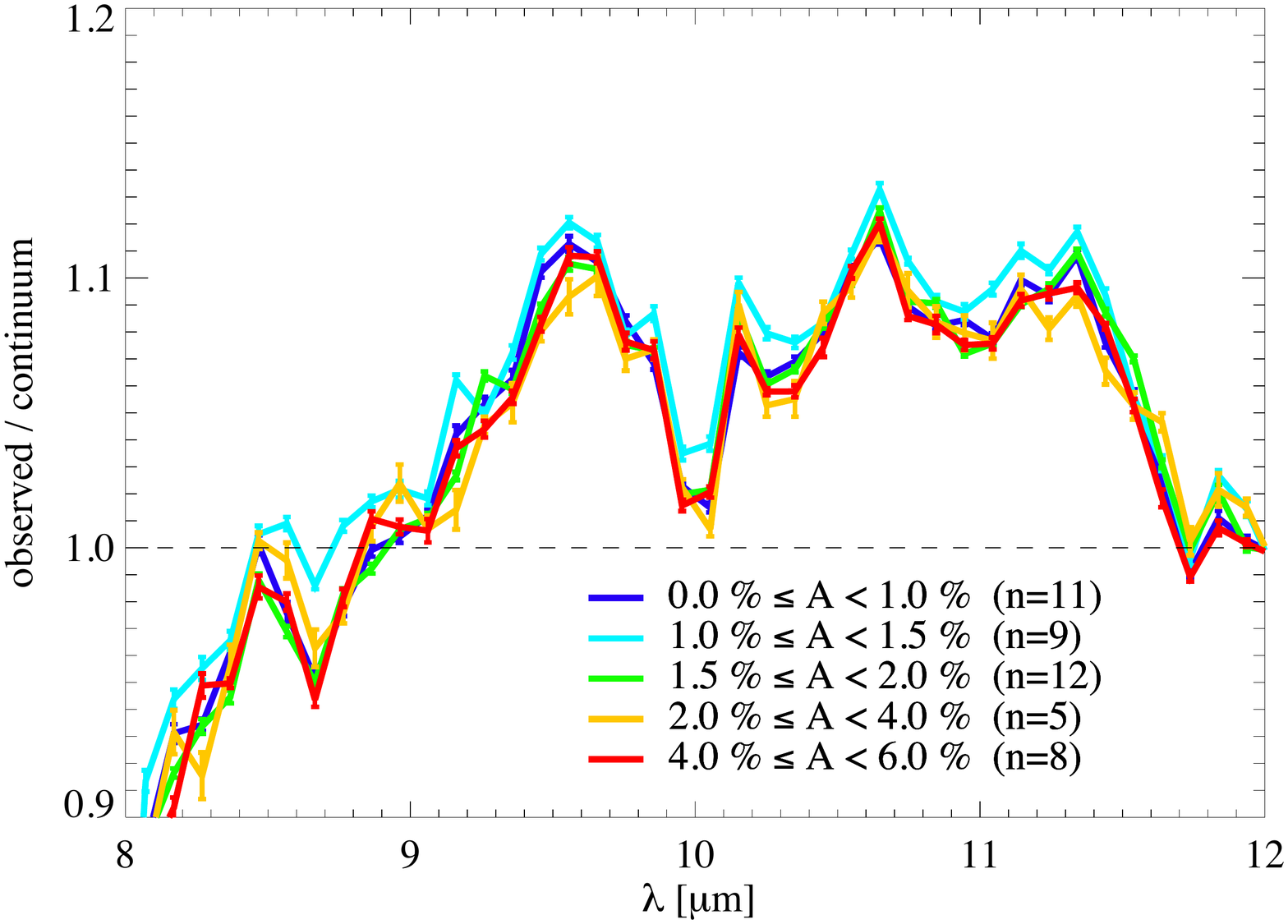}
\end{center}
\caption{The observed/continuum spectra averaged over each $A$-bin. The error bars represent the standard errors in the averaging. We show the number n of data points in each bin in the legend.}
\label{ast_compare_fig}
\end{figure}

\begin{figure}[htbp]
\begin{center}
     \includegraphics[width=\linewidth]{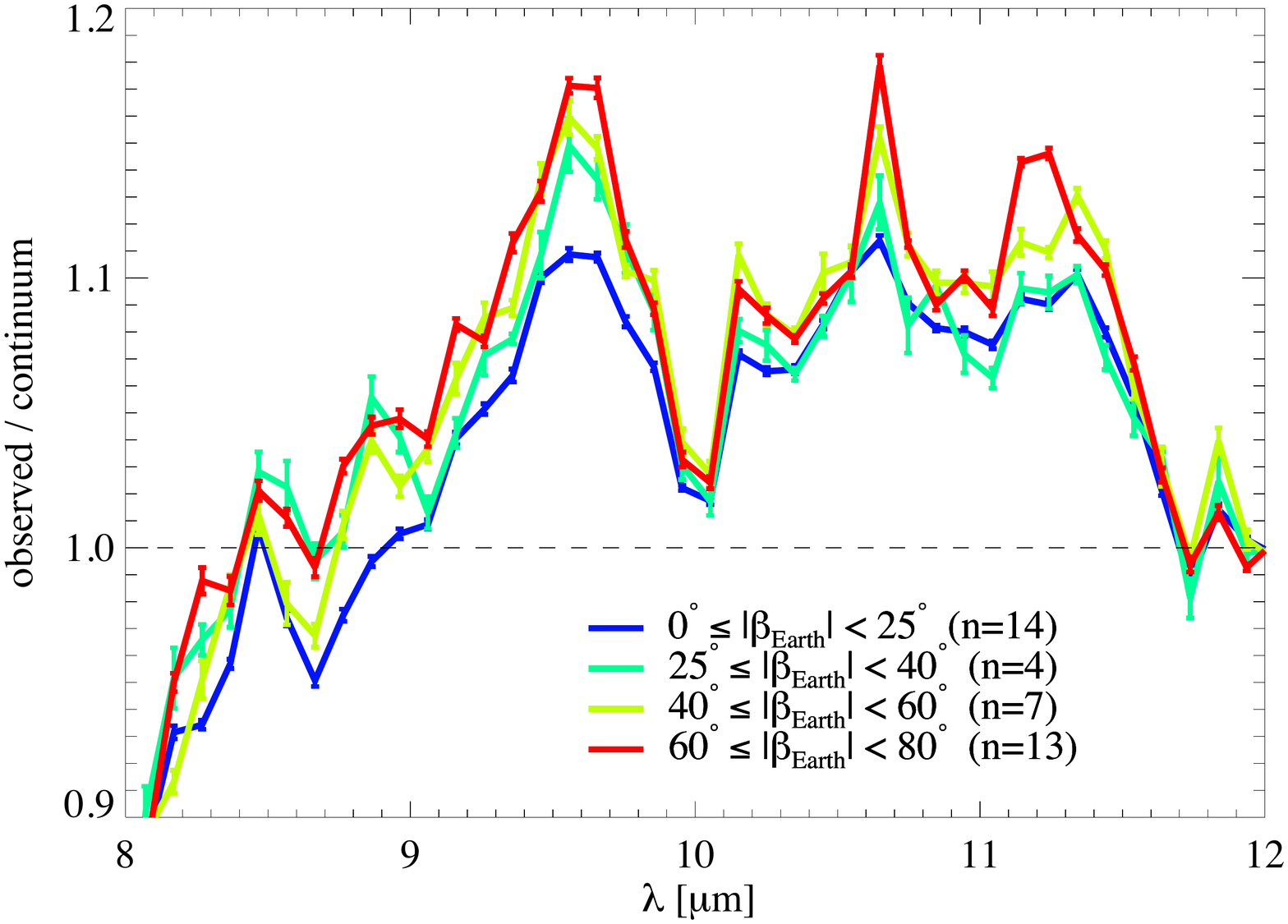}
\end{center}
\caption{The observed/continuum spectra averaged over each $|\beta_{\oplus}|$-bin. The error bars represent the standard errors in the averaging. We show the number n of data points in each bin in the legend.}
\label{elat_compare_fig}
\end{figure}

As another indicator of the sky direction, we considered the ecliptic latitudes $\beta_{\oplus}$ of the pointing directions. 
Figure~\ref{2d_hist_fig} is a two-dimensional histogram of $A$ vs. $|\beta_{\oplus}|$ for all the observed directions. 
We divided the dataset in the directions with $|\beta_{\oplus}| < 20^{\circ}$ into five $A$-bins (in percentages): $0.0 \leq A < 1.0$, $1.0 \leq A < 1.5$, $1.5 \leq A < 2.0$, $2.0 \leq A < 4.0$, and $4.0 \leq A < 6.0$. 
The observed/continuum spectra averaged over each $A$-bin are presented in Figure~\ref{ast_compare_fig}, and we compare the feature shapes among them. 
Similarly, we divided the dataset with $A < 1$\% into four $|\beta_{\oplus}|$-bins: $0^{\circ} \leq |\beta_{\oplus}| < 25^{\circ}$, $25^{\circ} \leq |\beta_{\oplus}| < 40^{\circ}$, $40^{\circ} \leq |\beta_{\oplus}| < 60^{\circ}$, and $60^{\circ} \leq |\beta_{\oplus}| < 80^{\circ}$, excepting the data at $|\beta_{\oplus}| = 90^{\circ}$, where the number of data points was too small. 
Figure~\ref{elat_compare_fig} shows the observed/continuum spectra averaged over each $|\beta_{\oplus}|$-bin, and we discuss below the differences in the feature shapes. 

As described in subsection \ref{averaged_feature_sec}, the $Q_\mathrm{abs}$ values of the current initial candidates cannot precisely reproduce the peak-wavelength positions in the observed spectra. 
We therefore cannot examine the grain properties by fitting the superposition of these $Q_\mathrm{abs}$ values to the observed spectra. 
Instead, for quantitative determinations of the feature shapes, we define the following four parameters: the equivalent width $EW_\mathrm{whole}$ of the whole excess emission, the ratio $EW_\mathrm{10.0-10.5}/EW_\mathrm{9.0-9.5}$ of the equivalent widths at 10.0--10.5~$\mu$m and 9.0--9.5~$\mu$m, the wavelength shift $\Delta \lambda_\mathrm{peak}$ of the peak excesses due to the crystalline silicates, and the equivalent width ratio $EW_\mathrm{fo}/EW_\mathrm{en}$ of the peak excesses due to the crystalline silicates. 
We remind that small grains ($\leq$ 1~$\mu$m) cause excess emission and contribute to these parameters. 
It means the first parameter $EW_\mathrm{whole}$ is related to the fraction of small grains in the line of sight and the other three parameters can be used for the investigation of the properties of small IPD grains. 
We describe below the detailed definition of each parameter and its dependence on $A$ and $|\beta_{\oplus}|$.

\subsubsection{Equivalent width of the whole excess emission}

\begin{figure}[htbp]
\begin{center}
     \includegraphics[width=\linewidth]{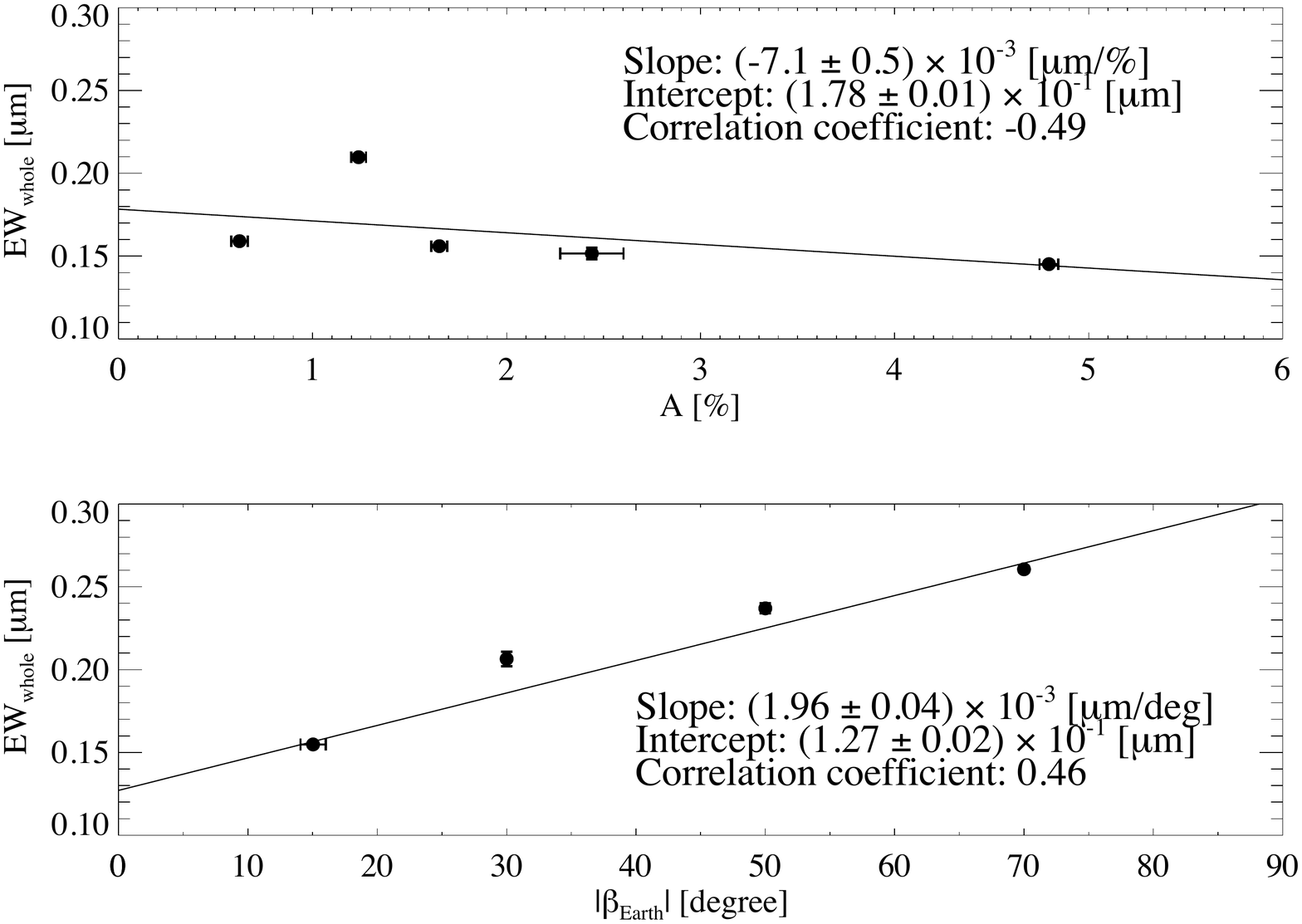}
\end{center}
\caption{The dependence of the equivalent width of the whole excess emission ($EW_\mathrm{whole}$) on the brightness contribution from the dust bands ($A$; the top panel) and on the ecliptic latitude ($|\beta_{\oplus}|$; the bottom panel). Each filled circle is calculated from the observed/continuum spectrum averaged over each bin and shown in Figures \ref{ast_compare_fig} and \ref{elat_compare_fig}. The position on the horizontal axis corresponds to the mean value in the bin. The solid lines are the linear functions that best fit the dependence, and the results of the fits are summarized in Table~\ref{linear_fit_parm}.}
\label{EWwhole_fig}
\end{figure}

The first parameter, the equivalent width of the whole excess emission, is defined by the equation: 
\begin{equation}
EW_\mathrm{whole} = \int_{8 \mathrm{\mu m}}^{12 \mathrm{\mu m}} \frac{Z_{\lambda} - C_{\lambda}}{C_{\lambda}} d\lambda ~~~[\mathrm{\mu m}]~. 
\label{EWwhole_eq}
\end{equation}
Here $Z_{\lambda}$ and $C_{\lambda}$ are the zodiacal emission spectrum given by equation (\ref{ZE_eq}) and the continuum spectrum given by equation (\ref{continuum_eq}), respectively. 
This represents the excess strength in the 8--12~$\mu$m range. 
We calculated this quantity for the observed/continuum spectra averaged over each bin, and we plot the dependence on $A$ and $|\beta_{\oplus}|$ in Figure~\ref{EWwhole_fig}. 
The parameter $EW_\mathrm{whole}$ seems to be negatively correlated with $A$, although it is not well-expressed by a linear function, according to the extremely large reduced-$\chi^{2}$ value of the linear fit. 
On the other hand, $EW_\mathrm{whole}$ does have a clear positive correlation with $|\beta_{\oplus}|$. 
This indicates that the IPD at higher $|\beta_{\oplus}|$ (or higher $A$) shows stronger (weaker) excess emission in the 8--12~$\mu$m range. 

The derived excess strengths and the $|\beta_{\oplus}|$-dependence are roughly consistent with a previous study by Reach~et~al.~(2003)~\cite{Reach_2003} of excess strengths at some $\beta_{\oplus}$ and some solar elongations. 
This variation of $EW_\mathrm{whole}$ can be interpreted as a difference in the IPD grain-size frequency distribution along the lines of sight in different directions. 
Since an increase in the fraction of small grains seems to strengthen excess emission according to Figure~\ref{Qabs_fig}, we can say that lines of sight at higher $|\beta_{\oplus}|$ include a more fraction of small grains, while the fraction of small grains exhibits a relative decrease in directions toward the dust bands. 

\subsubsection{Ratio of equivalent widths at 10.0--10.5~$\mu$m and 9.0--9.5~$\mu$m}

\begin{figure}[htbp]
\begin{center}
     \includegraphics[width=\linewidth]{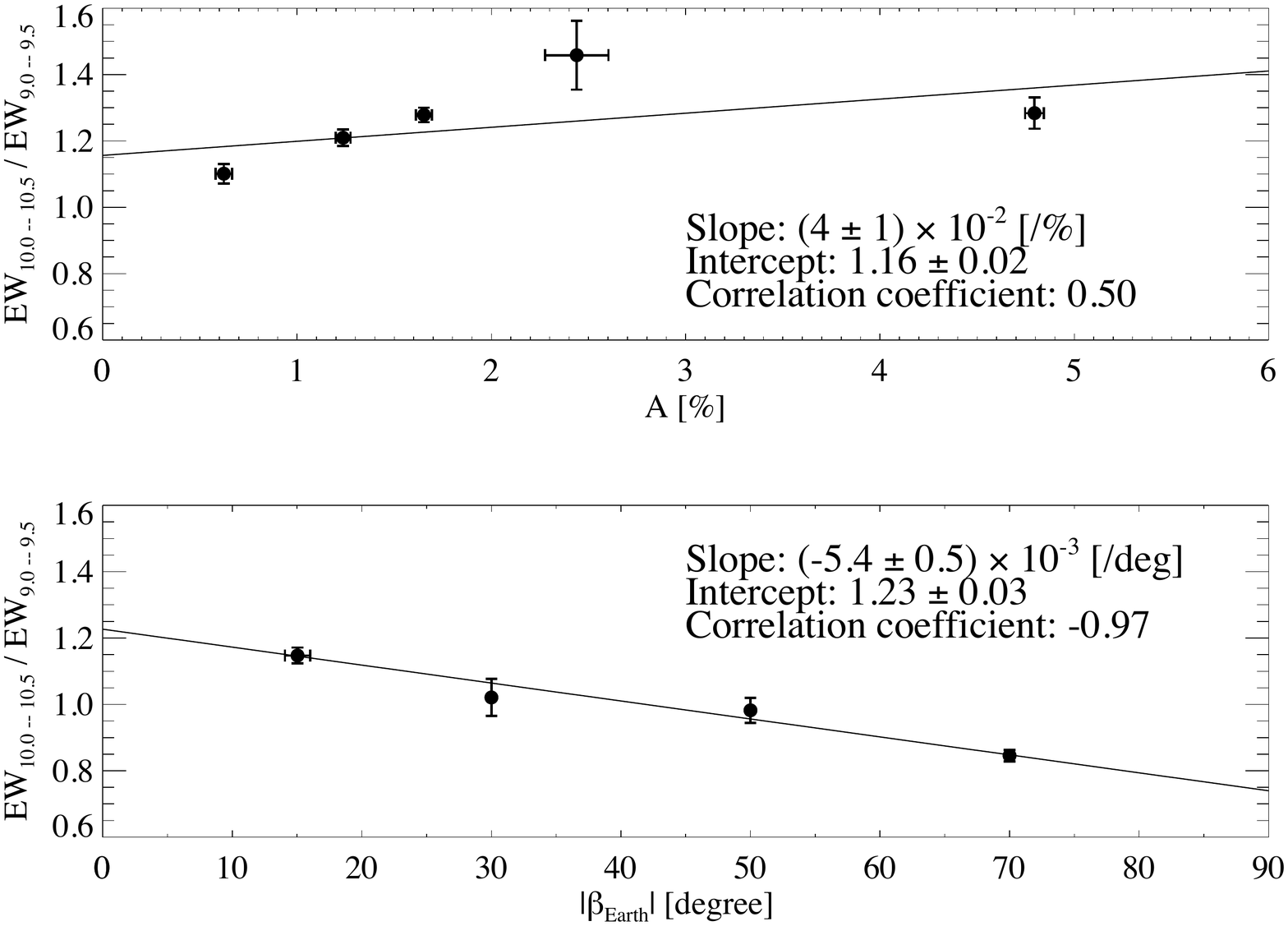}
\end{center}
\caption{Dependence of the equivalent width ratio in 10.0 -- 10.5~$\mu$m and 9.0 -- 9.5~$\mu$m ($EW_\mathrm{10.0-10.5}/EW_\mathrm{9.0-9.5}$) on the brightness contribution from the dust bands ($A$; the top panel) and on the ecliptic latitude ($|\beta_{\oplus}|$; the bottom panel). Other explanations are the same as in Figure~\ref{EWwhole_fig}. The errors in the y-axis at $A$=2.5 \% and $|\beta_{\oplus}|$=30$^{\circ}$ and 50$^{\circ}$ were relatively large because of the small numbers of data points.}
\label{EWolipyr_fig}
\end{figure}

According to Figure~\ref{Qabs_fig}, small grains ($\leq$ 1~$\mu$m) of amorphous olivine and amorphous pyroxene produce a smooth convex excess around 10~$\mu$m and 9~$\mu$m, respectively. 
As a tentative indicator of the olivine/(olivine+pyroxene) ratio in small amorphous grains, we define the ratio of equivalent widths in the two wavelength regions, 10.0--10.5~$\mu$m and 9.0--9.5~$\mu$m: 
\begin{equation}
EW_\mathrm{10.0-10.5}/EW_\mathrm{9.0-9.5} = \int_{10.0 \mathrm{\mu m}}^{10.5 \mathrm{\mu m}} \frac{Z_{\lambda} - C_{\lambda}}{C_{\lambda}} d\lambda \bigg/ \int_{9.0 \mathrm{\mu m}}^{9.5 \mathrm{\mu m}} \frac{Z_{\lambda} - C_{\lambda}}{C_{\lambda}} d\lambda ~.
\label{EWolipyr_eq}
\end{equation}
We selected these wavelength regions to avoid contamination by the peaks due to small crystalline silicates as much as possible. 
The resulting correlation with $A$ and $|\beta_{\oplus}|$ is shown in Figure~\ref{EWolipyr_fig}. 
The ratio $EW_\mathrm{10.0-10.5}/EW_\mathrm{9.0-9.5}$ shows a positive (negative) correlation with $A$ ($|\beta_{\oplus}|$). 
This trend means that small grains in the dust bands shows relatively strong olivine-like features, while the small IPD grains at high $|\beta_{\oplus}|$ appears to have more pyroxene-like features than typical small IPD grains do. 

\subsubsection{Wavelength shift of the peak excesses due to small crystalline silicates}

The two remaining parameters are focused on more detailed feature shapes, which the AKARI observations have enabled us to examine for the first time. 
We investigated the wavelength shifts and equivalent-width ratios of the peak excesses due to the small crystalline silicates: forsterite and enstatite. 
In order to compare such parameters among peaks from the same vibration modes in different directions, we need first to identify which peaks originate from the same vibration mode. 
We established a reference wavelength $\lambda_\mathrm{ref}$ for each main peak that is common to all the observing directions, and we assumed that the peaks seen within the two neighboring wavelength bins around $\lambda_\mathrm{ref}$ (corresponding roughly to the region $\lambda_\mathrm{ref} \pm$0.25~$\mu$m) originate from the same vibration mode. 

\begin{figure}[htbp]
\begin{center}
     \includegraphics[width=0.85\linewidth]{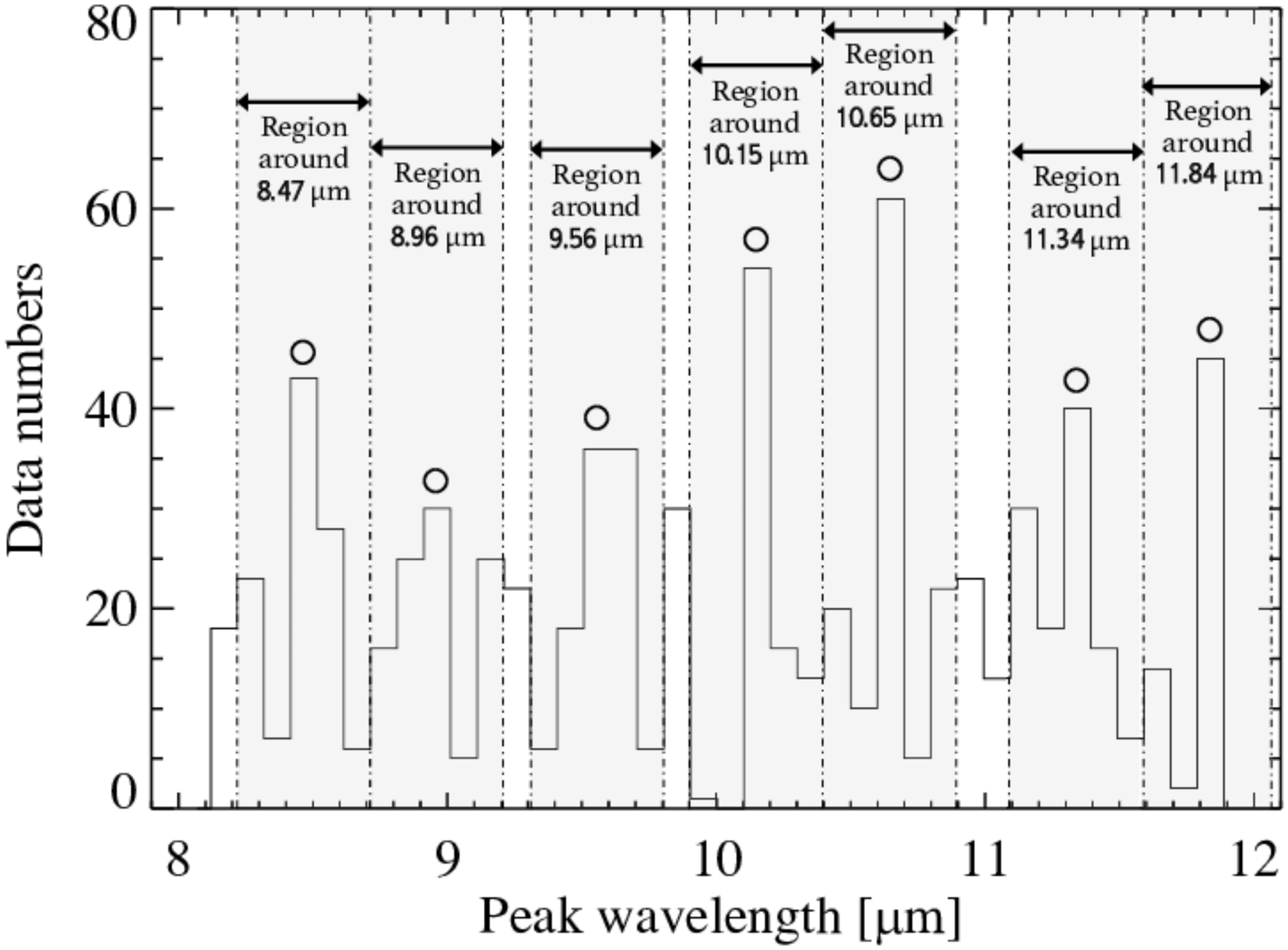}
\end{center}
\caption{The histogram of peak wavelengths. Each wavelength bin has a width of $\sim$0.1~$\mu$m. Wavelength bins selected to serve as reference wavelengths are marked with a circle at the top. We used the data in the seven peak regions indicated by the arrows.}
\label{peakWVhist_fig}
\end{figure}

The procedure we used to establish $\lambda_\mathrm{ref}$ is the following. 
\begin{enumerate}
\item For the observed/continuum spectra in all 74 observations, we searched wavelength bins with a peak, which show a higher observed/continuum ratio than the neighboring bins. 
\item For all wavelength bins from 8 to 12~$\mu$m, we counted the number of data points that have a peak at the bin. The resulting histogram is shown in Figure~\ref{peakWVhist_fig}. 
\item We selected wavelength bins which have a peak most frequently among the five successive bins around it, and we defined the center wavelengths of the selected bins as $\lambda_\mathrm{ref}$. Around 9.6~$\mu$m, however, two neighboring wavelength bins had the same number of data points. For this case, we chose the center wavelength of the bin at the shorter wavelength as $\lambda_\mathrm{ref}$. 
\end{enumerate}
In this way, we found seven peaks that are commonly seen in many directions, and we fixed the corresponding reference wavelengths at $\lambda_\mathrm{ref}$ = 8.47, 8.96, 9.56, 10.15, 10.65, 11.34, and 11.84~$\mu$m. 

\begin{figure}[htbp]
\begin{center}
     \includegraphics[width=0.95\linewidth]{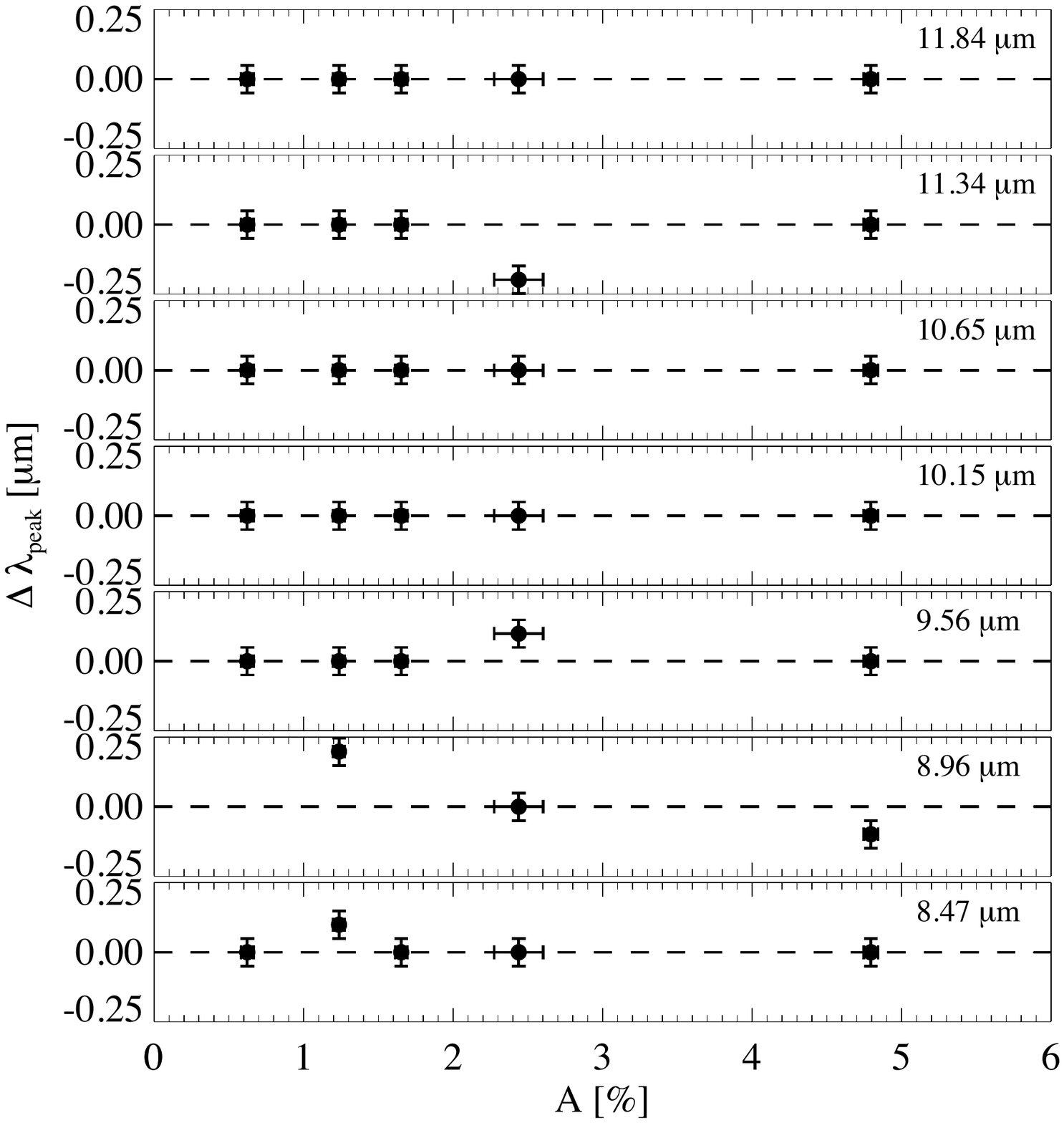}
     \includegraphics[width=0.95\linewidth]{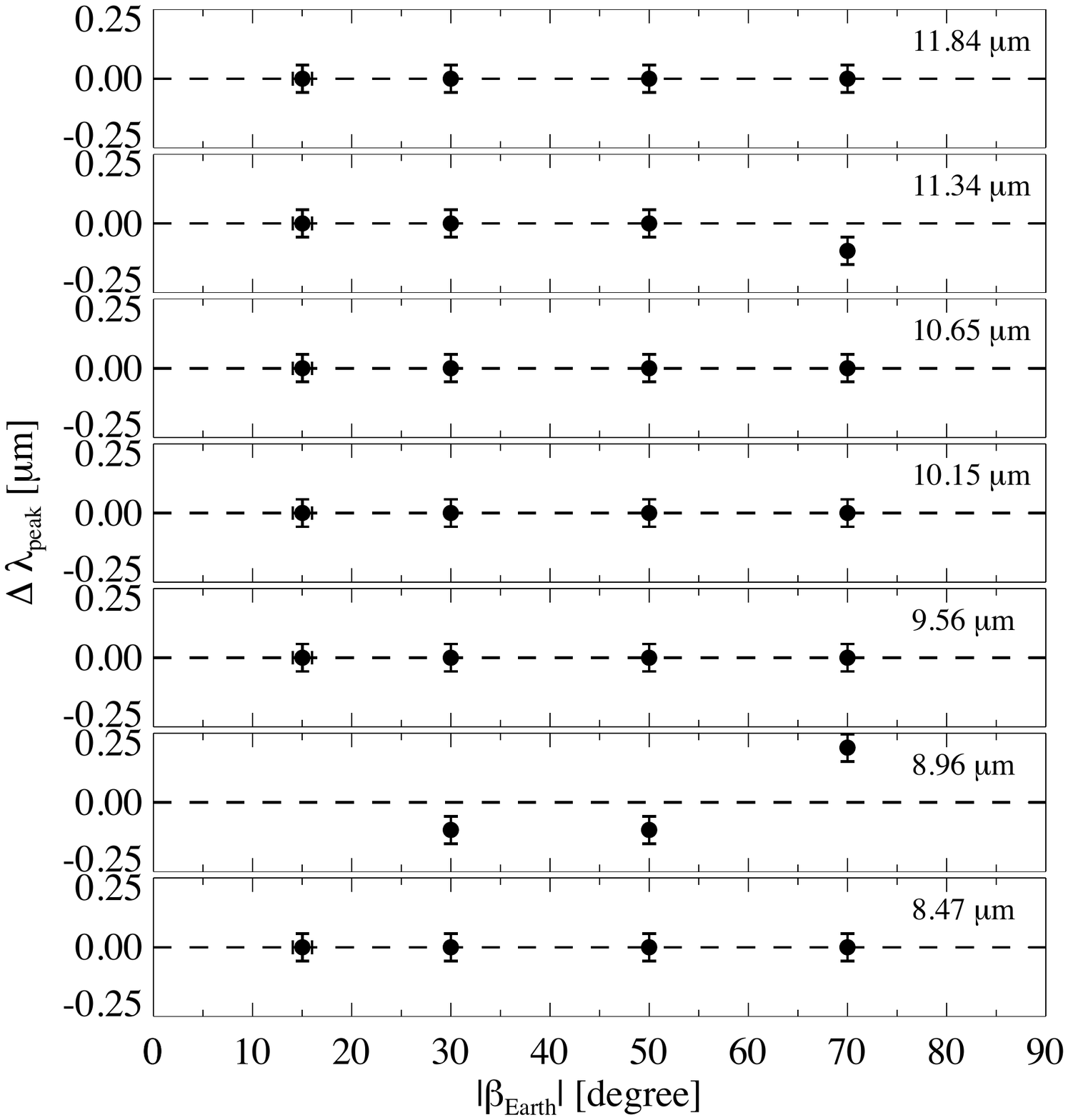}
\end{center}
\caption{Dependence of the wavelength shift $\Delta \lambda_\mathrm{peak}$ of the peak excess due to small crystalline silicates on the brightness contribution from the dust bands ($A$; the top panels) and on the ecliptic latitude ($|\beta_{\oplus}|$; the bottom panels). Each panel indicates the result for the peak in the region around $\lambda_\mathrm{ref}$, which is noted at the upper right side of the panel and corresponds to the dotted line. Other explanations are the same as in Figure~\ref{EWwhole_fig}. In the region around 8.96~$\mu$m, some average observed/continuum spectra  do not show a peak structure.} 
\label{WS_fig}
\end{figure}

The position of a peak wavelength for each of the 74 observations is not always identical to $\lambda_\mathrm{ref}$. 
From each of the seven regions of $\lambda_\mathrm{ref} \pm 2$ wavelength bins in each observed/continuum spectrum, we selected the wavelength bin that shows the most significant peak. 
We defined the center wavelength of that bin as the peak wavelength in each individual spectrum, denoting it by $\lambda_\mathrm{peak}$. 
The peak-wavelength shift in the region around $\lambda_\mathrm{ref}$ is then given by 
\begin{equation}
\Delta \lambda_\mathrm{peak} = \lambda_\mathrm{peak} - \lambda_\mathrm{ref}  ~~~[\mathrm{\mu m}].
\label{WS_eq}
\end{equation}
Although the absolute values of these quantities do not have strong meanings because it depends on the definition, the relative variations of $\Delta \lambda_\mathrm{peak}$ among the different sky directions imply differences in the grain properties, which affect the wavelength positions of their emission peaks. 
Figure~\ref{WS_fig} shows the resulting correlation of $\Delta \lambda_\mathrm{peak}$ in each peak region with $A$ and $|\beta_{\oplus}|$. 
$\Delta \lambda_\mathrm{peak}$ seems not to change in most regions, except for the region around 11.34~$\mu$m and 8.96~$\mu$m. 
This means that the grain properties such as metal composition and crystal morphology do not dramatically change depending on the sky direction (see section~\ref{averaged_feature_sec}). 
The region around 11.34~$\mu$m is too complicated to be discussed deeply because two peaks originating from both enstatite and forsterite contaminate this region. 

\subsubsection{Equivalent-width ratio of peak excesses due to small crystalline silicates}

\begin{figure}[htbp]
\begin{center}
     \includegraphics[width=\linewidth]{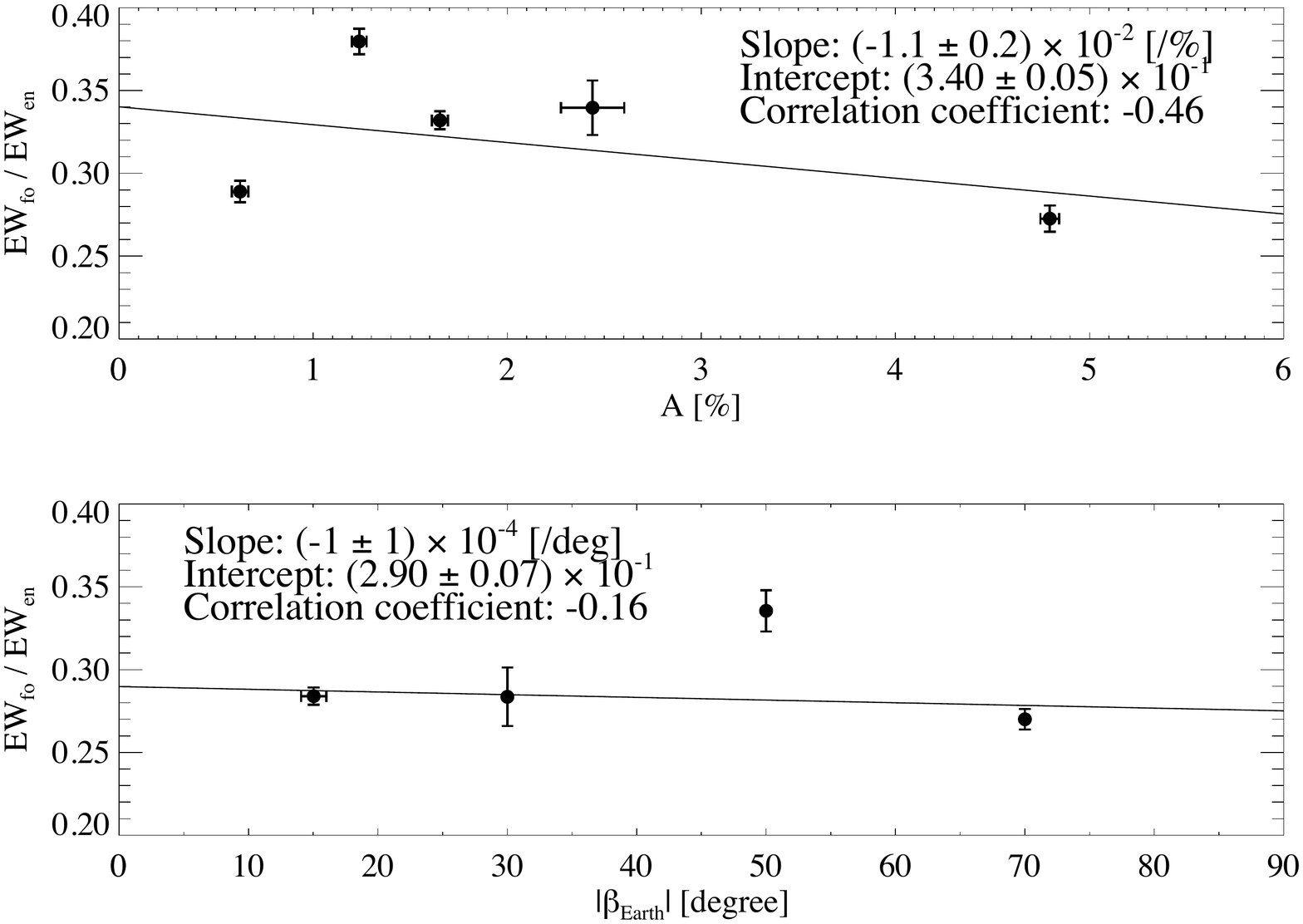}
\end{center}
\caption{Dependence of the equivalent-width ratio $EW_\mathrm{fo}/EW_\mathrm{en}$ of the peak excesses due to small crystalline silicates on the brightness contribution from the dust bands ($A$; the top panel) and on the ecliptic latitude ($|\beta_{\oplus}|$; the bottom panel). Other explanations are the same as in Figure~\ref{EWwhole_fig}. The errors in the y-direction at $A$=2.5\% and $|\beta_{\oplus}|$=30$^{\circ}$, and 50$^{\circ}$ are relatively large because of the small numbers of data points.}
\label{EWfoen_fig}
\end{figure}

The last feature-shape parameter we consider is the equivalent-width ratio, $EW_\mathrm{fo}/EW_\mathrm{en}$, of the peak excesses due to small forsterite and enstatite grains. 
We took into account the peaks around 10.15 and 11.84~$\mu$m to calculate the equivalent width of the forsterite peak, while the peak around 9.56 and 10.65~$\mu$m to calculate the equivalent width of the enstatite peak. 
We thus defined this parameter as
\begin{equation}
\frac{EW_\mathrm{fo}}{EW_\mathrm{en}} = \frac{EW_\mathrm{peak}(10.15~\mathrm{\mu m}) + EW_\mathrm{peak}(11.84~\mathrm{\mu m})}{EW_\mathrm{peak}(9.56~\mathrm{\mu m}) + EW_\mathrm{peak}(10.65~\mathrm{\mu m})} ~, 
\label{EWfoen_eq}
\end{equation}
where
\begin{equation}
EW_\mathrm{peak}(\lambda_\mathrm{ref}) = \int_{\lambda_\mathrm{peak}(\lambda_\mathrm{ref}) - 0.25 \mathrm{\mu m}}^{\lambda_\mathrm{peak}(\lambda_\mathrm{ref}) + 0.25 \mathrm{\mu m}} \frac{Z_{\lambda} - C_{\lambda}}{C_{\lambda}} d\lambda ~~~[\mathrm{\mu m}]~. 
\label{EWfoen2_eq}
\end{equation}
Here, $\lambda_\mathrm{peak}(\lambda_\mathrm{ref})$ is the peak wavelength in the region around $\lambda_\mathrm{ref}$ at the individual $A$- or $|\beta_{\oplus}|$-bin. 
This means that the wavelength intervals of these integrations change depending upon the wavelength shift in each bin. 
As shown in Figure~\ref{EWfoen_fig}, $EW_\mathrm{fo}/EW_\mathrm{en}$ exhibits a negative correlation with $A$, although the dispersion is significant. 
We could not find any correlation between $EW_\mathrm{fo}/EW_\mathrm{en}$ and $|\beta_{\oplus}|$ at least in this statistical processing, regarding the low correlation coefficient. 

\begin{table}[htbp]
  \centering
  \caption{The slopes of the linear functions of $A$ or $|\beta_{\oplus}|$ that best fit $EW_\mathrm{whole}$, $EW_\mathrm{10.0--10.5}/EW_\mathrm{9.0--9.5}$, and $EW_\mathrm{fo}/EW_\mathrm{en}$. The reduced-$\chi^{2}$ of each fit and correlation coefficient $r$ are also shown. These slope values are plotted in Figures \ref{EWwhole_fig}, \ref{EWolipyr_fig}, and \ref{EWfoen_fig}, respectively.}
  \label{linear_fit_parm}
  \begin{threeparttable}
  \begin{tabular}{|c|c|c|c|c|c|c|} \hline
    parameter  & \multicolumn{3}{|c|}{$A$-dependence} & \multicolumn{3}{|c|}{$|\beta_{\oplus}|$-dependence} \\ \cline{2-7}
      type     & Slope~\tnote{1}  & reduced-$\chi^{2}$ & $r$ & Slope~\tnote{2} & reduced-$\chi^{2}$ & $r$ \\ \hline \hline
            $EW_\mathrm{whole}$                  &  $(-7.1 \pm 0.5) \times 10^{-3}$ &  287.42 & -0.49 &  $(1.96 \pm 0.04) \times 10^{-3}$  &  21.68 & 0.46 \\ \hline
  $EW_\mathrm{10.0-10.5}/EW_\mathrm{9.0-9.5}$    &  $(4 \pm 1) \times 10^{-2}$ &  6.44 & 0.50 & $(-5.4 \pm 0.5) \times 10^{-3}$  &  0.56  & -0.97 \\ \hline
         $EW_\mathrm{fo}/EW_\mathrm{en}$         &  $(-1.0 \pm 0.2) \times 10^{-2}$ &  34.19 & -0.46 & $(-1 \pm 1) \times 10^{-4}$  &  10.42 & -0.16 \\ \hline
  \end{tabular}

 \begin{tablenotes}\footnotesize
 \item[1] Units of [$\mu$m/\%] for $EW_\mathrm{whole}$, [/\%] for $EW_\mathrm{10.0-10.5}/EW_\mathrm{9.0-9.5}$ and $EW_\mathrm{fo}/EW_\mathrm{en}$. The maximum amplitude of $A$ is $<$ 6 \%. \\
 \item[2] Units of [$\mu$m/deg] for $EW_\mathrm{whole}$, [/deg] for $EW_\mathrm{10.0-10.5}/EW_\mathrm{9.0-9.5}$ and $EW_\mathrm{fo}/EW_\mathrm{en}$. The possible maximum amplitude of $|\beta_{\oplus}|$ is 90 degree. 
 \end{tablenotes}
  \end{threeparttable}
\end{table}

\section{Discussion}

In section \ref{result_sec}, we presented the typical properties of the small IPD grains determined from the average feature shapes and we described the variations of the feature shapes and the related grain properties among the different sky directions. 
We next describe the implications of these results. 

\subsection{Implications of the typical properties}

According to a comparison between the feature shapes averaged over all directions and the absorption coefficients of candidate minerals, we found that the IPD typically includes small silicate crystals such as enstatite. 

The most significant source of the IPD around 1 au is thought to be comets~\cite{Nesvorny_2010} and some of the cometary IPD particles collected in the stratosphere show clear enstatite peaks~\cite{Merouane_2014}. 
The results we have obtained thus seem consistent with previous work.  

In the interstellar medium, the degree of crystallinity of silicate dust is less than 2 \%~\cite{Kemper_2005}. 
Therefore, the silicate crystals seen in the IPD are likely to have been formed by re-condensation from gas in the solar nebula and/or by annealing of amorphous interstellar (i.e., pre-solar) dust. 
In the pressure environment at the mid-plane of the proto-solar disk, the equilibrium condensation temperatures of enstatite and forsterite are 1300 K and 1400 K, respectively~\cite{Gail_2004}. 
Hallenbeck~et~al.~(2000)~\cite{Hallenbeck_2000} have shown that the annealing temperature of silicates is about 950 K. 
However, such high temperature is possible only near the Sun (at least less than a few au), if we assume the global radial dependence of the disk temperature in a stationary disk model~\cite{Gail_2001, Bell_1997, Murata_2009}. 
This indicates radial mixing of the dust population and/or local heating in the proto-solar disk. 


\subsection{Implications of the variation with $A$}

By comparing the feature shapes among different sky directions with the contribution, $A$, from the dust bands, we found several correlations between the feature shapes and $A$. 
This may imply differing grain properties of the IPD originating from asteroids and comets. 
Sykes \& Greenberg (1986)~\cite{Sykes_1986} and Nesvorn\'{y}~et~al.~(2003)~\cite{Nesvorny_2003} propose that the dust bands were formed by the IPD from asteroids, while the main source of other IPD components is comets. 

According to the negative correlation between $EW_\mathrm{whole}$ and $A$ shown in Figure~\ref{EWwhole_fig}, the asteroidal IPD may have a size frequency distribution biased toward large grains. 
It may be because the dust continuously supplied by collisional cascades among asteroids can include a lot of large grains during the cascades, while the cometary IPD is ejected from melting icy mantles mainly as small grains~\cite{Shinnaka_2018}. 
Thus, differences in the supply processes may cause the different size frequency distributions. 

Other differences in the properties of the asteroidal and cometary IPD grains have been found in previous work. 
A unique property of the asteroidal IPD is hydration. 
Schramm~et~al.~(1989)~\cite{Schramm_1989} and Germani~et~al.~(1990)~\cite{Germani_1990} found hydrated minerals in the IPD particles thought to come from asteroids, while the cometary IPD seems to be anhydrous. 
The CM chondrites, which are a type of carbonaceous meteorite thought to originate from C-type asteroids, are also well-known to contain hydrated materials~\cite{McSween_1979, McSween_1987}. 
In addition, Usui~et~al.~(2018)~\cite{Usui_2018} discovered hydrated minerals in most C-complex asteroids. 
Common hydrated minerals are phyllosilicates like serpentine and talc, which are produced by aqueous alteration from forsterite or enstatite~\cite{Ganguly_1995, Bischoff_1998, Tomeoka_1989}. 
They show a smooth excess feature with a peak around 10~$\mu$m, similar to that of amorphous olivine~\cite{Beck_2014}. 
We suggest the possibility that the feature caused by phyllosilicate may be responsible for increasing the ratio $EW_\mathrm{10.0-10.5}/EW_\mathrm{9.0-9.5}$ at the directions toward the dust bands, as shown in Figure~\ref{EWolipyr_fig}, without an increase in the fraction of amorphous olivine. 
In such a case, the peaks originating from both forsterite and enstatite will appear to be reduced, because aqueous alteration changes the forsterite and enstatite into serpentine and talc, respectively. 
Considering that the phyllosilicate peak can replenish the excess strength around 10~$\mu$m but not in the olivine peak around 11.84~$\mu$m, this scenario may also explain the negative correlation between $A$ and $EW_\mathrm{fo}/EW_\mathrm{en}$ shown in Figure~\ref{EWfoen_fig}. 
Since the most significant dust band at $\beta_{\oplus}=\pm 9^{\circ}.3$~\cite{Kelsall_1998} is thought to be formed by the IPD from the C-type Veritas family~\cite{Nesvorny_2003}, this scenario is consistent with the result of previous work which found hydration in C-type asteroids and CM chondrites.

\subsection{Implication of the variation with $|\beta_{\oplus}|$}

As mentioned in section \ref{direction_compare_sec}, the $|\beta_{\oplus}|$-dependence of $EW_\mathrm{whole}$, the equivalent width of the whole excess emission in 8--12~$\mu$m band, indicates an increase in the fraction of small grains at higher $|\beta_{\oplus}|$. 
This can be explained by the radial dependence of the size frequency distribution of the IPD grains. 
Jehn~(2000)~\cite{Jehn_2000} found that large grains exist more frequently at a few au from the Sun as compared with near the Earth, on the ecliptic plane where the IPD distributes convergently. 
Since the IPD at radial distances far from the Earth is included only in lines of sight at low $|\beta_{\oplus}|$, the size frequency distribution of the IPD grains at high $|\beta_{\oplus}|$ is relatively biased toward small grains. 
The radial dependence of the size frequency distribution is thought to be caused by the collfisional cascade during the accretion of IPD particles toward the Sun owing to Poynting-Robertson drag~\cite{Grun_1985}. 

We also found differences in the mineral composition of the small IPD grains as a function of $|\beta_{\oplus}|$. 
This may imply differences in the grain properties of the IPD originating from different types of comets. 
According to Nesvorn\'{y}~et~al.~(2010)~\cite{Nesvorny_2010}, the IPD from the Jupiter Family Comets (JFCs) is the most significant and is distributed mainly around the ecliptic plane, barely spreading toward the ecliptic poles. 
On the other hand, the IPD from Oort Cloud Comets (OCCs) is known to have an isotropic distribution, owing to the wide range of inclinations of the OCCs. 
This means that the fraction of the IPD originating from the OCCs becomes relatively larger at higher $|\beta_{\oplus}|$. 
One of reasons for the negative $|\beta_{\oplus}|$-dependence of $EW_\mathrm{10.0-10.5}/EW_\mathrm{9.0-9.5}$ may be because the IPD from the OCCs has a lower olivine/(olivine+pyroxene) ratio than the IPD from the JFCs regarding at least small amorphous grains. 
Such differences indicate the different forming regions of the JFCs and the OCCs.

\section{Conclusion}

Using mid-infrared slit-spectroscopic data of the zodiacal emission obtained with AKARI/IRC, we have succeeded in detecting details of the shapes of the excess emission features in the 9--12~$\mu$m range. 
The feature shape averaged over all directions indicates that the IPD typically includes small silicate crystals such as enstatite, suggesting the existence of radial mixing and/or local heating of the dust in the proto-solar disk. 
We also found variations in the feature shapes among different sky directions. 
From investigations of $EW_\mathrm{whole}$, we found that the spectra at higher $|\beta_{\oplus}|$ showed a stronger excess, which indicates an increase in the fraction of small grains included in the line of sight at higher ecliptic latitudes. 
On the other hand, the negative correlation with $A$ indicates that the size frequency distribution of the IPD in the dust bands is biased toward relatively large grains. 
The positive and negative correlations with $A$ of $EW_\mathrm{10.0-10.5}/EW_\mathrm{9.0-9.5}$ and $EW_\mathrm{fo}/EW_\mathrm{en}$, respectively, can be qualitatively explained by aqueous alteration from forsterite and enstatite to phyllosilicates in the asteroidal IPD. 
From the negative $|\beta_{\oplus}|$-dependence of $EW_\mathrm{10.0-10.5}/EW_\mathrm{9.0-9.5}$, we found the possibility that the IPD at higher ecliptic latitudes may have lower olivine/(olivine+pyroxene) ratio for small amorphous grains. 
Such variation of the mineral composition of the IPD at different ecliptic latitudes may imply the difference of the mineral composition between the IPD from the JFCs and OCCs, because their distributions in ecliptic latitudes are different. 

\vspace*{5.0\baselineskip}

This research is based on observations with AKARI, a JAXA project with the participation of ESA. We would like to thank all the members of the AKARI project. 
This work was supported by two JSPS KAKENHI Grant-in-Aid for Scientific Research (C): Grant Number JP17K05381 and JP17K05636, and partly by the Astrobiology center and the National Astronomical Observatory of Japan. 
Since this paper is based on the PhD-thesis work of Aoi Takahashi, we greatly appreciate referees of the PhD-defense: Takahiro Iwata, Yuko Inatomi (ISAS/JAXA) and Hirokazu Kataza (University of Tokyo). 
Finally, we are sincerely thankful for pertinent comments from a reviewer
and editorial works. 


\clearpage

\appendix
\section{Elimination of artifacts}
\label{artifacts}

\begin{figure}[htbp]
\begin{center}
     \includegraphics[width=0.8\linewidth]{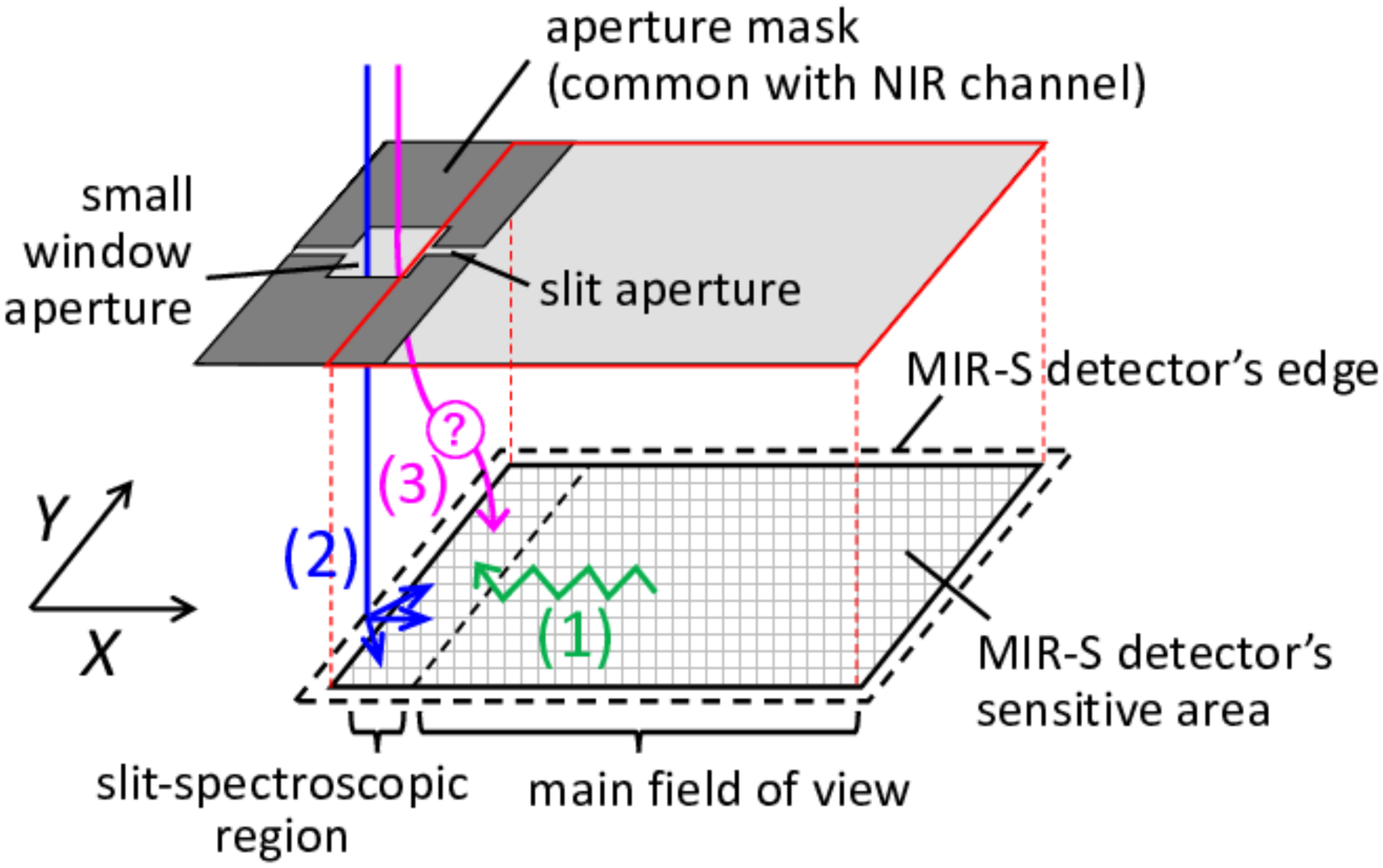}
\end{center}
\caption{A schematic view of three artificial components: (1) light scattered from detector pixels, (2) light scattered from the edge of the detector, and (3) ghost of a small aperture window. In actuality, some optical elements---like a beam splitter, two lenses, and a grism---are also included between the aperture mask and the detector. The $Y$ direction is the direction of wavelength dispersion.}
\label{artifact_schema_fig}
\end{figure}

\subsection{Light scattered from the detector pixels}

A small fraction of the light incident on a detector pixel is scattered into other pixels in the same row and column of the pixel array, even if the pixel is not saturated. 
From careful investigations of point-source imaging data obtained in the MIR-S channel, it has been empirically found that the light with intensity $S_{10}$ [ADU] incident at the pixel position ($X_{0}$, $Y_{0}$) is scattered into the pixel position ($X$, $Y_{0}$) with the fraction of 
\begin{equation}
G_{1}(X-X_{0}) = \frac{C}{1+(\frac{X-X_{0}}{22.6})^{2}} \mbox{~~~[ADU]}, 
\label{FOVscatter_eq}
\end{equation}
where $C = 7.73 \times 10^{-4}$ ($6.23 \times 10^{-4}$) for SG1 (SG2). 
This formula can be used also for the scattering along the Y-axis. 
The actual brightness of this component at each position is determined by the convolution of the light scattered from all pixels in the same row and column: 
\begin{equation}
S_{1}(X,Y) =  \int_{1}^{256} S_{10}(X_{0},Y) \times G_{1}(X-X_{0}) ~d X_{0} + \int_{1}^{256} S_{10}(X,Y_{0}) \times G_{1}(Y-Y_{0}) ~d Y_{0} \mbox{~~~[ADU]}. 
\label{FOVscatter_eq2}
\end{equation}
If we care about their effect on the slit-spectroscopic region, the light leaking from the main field of view along the X-axis is dominant [see (1) in Figure~\ref{artifact_schema_fig} and Appendix 2 of Sakon~et~al.~(2007)~\cite{Sakon_2007}]. 
We calculated the two-dimensional brightness distribution of the leaking light for each pointing data. 
After initial processing of the observed image (dark subtraction, linearity correction, and median determination), we assumed the brightness distribution in the processed image itself to be approximately the distribution of $S_{10}$, because the fraction of the scattered light that is included is as small as 10$^{-4}$ of the incident light [the order of magnitude of $C$ in equation (\ref{FOVscatter_eq})]. 
Using the distribution of $S_{10}$, we calculated the distribution of $S_{1}$ from equation (\ref{FOVscatter_eq2}). 
Panel (a) in Figure~\ref{artifact_fits_fig} shows an example of the estimated brightness distribution of this component. 
We subtracted it from the processed image. 

\subsection{Light scattered from the edge of the detector}

\begin{figure}[tb]
\begin{center}
     \includegraphics[width=0.85\linewidth]{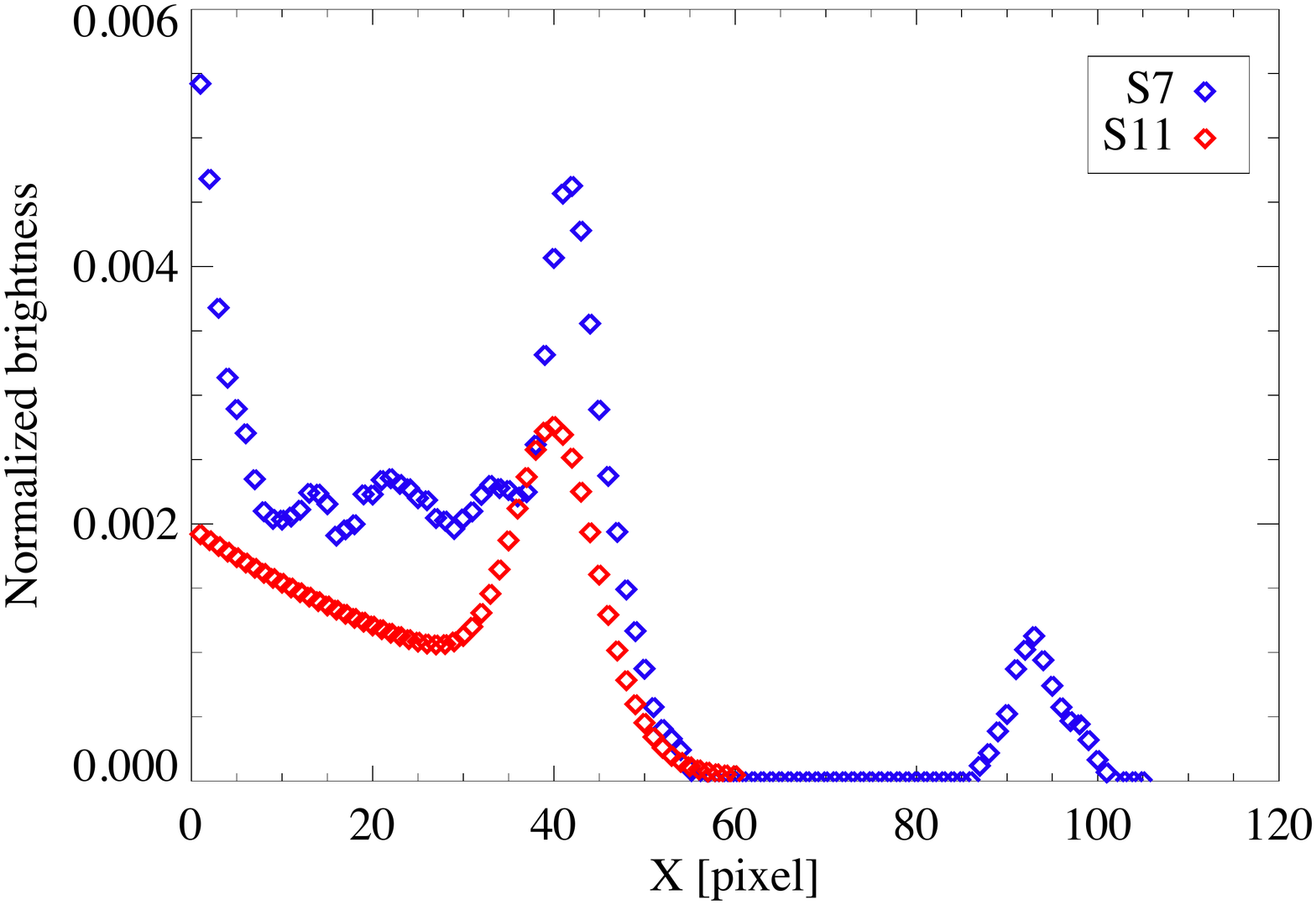}
\end{center}
\caption{The brightness profile of light scattered from the edge of the detector. The absolute value is normalized by the brightness of the source causing the scattered light. The filters S7 and S11 are used for the imaging observations, because their wavelength coverages correspond to those of SG1 and SG2, respectively. We used the profile obtained with S7 for SG1 and with S11 for SG2.}
\label{NPWscatter_prof_fig}
\end{figure}

The aperture mask we used was shared with the Near-Infrared (NIR) channel, and it had a small window for the spectroscopy of point sources in the NIR channel. 
Since the detector surface in the MIR-S channel is smaller than that in the NIR channel, some light that passes through this window is scattered by the edge of the detector in the MIR-S channel and contaminates the slit-spectroscopic region [see (2) in Figure~\ref{artifact_schema_fig}]. 
When we checked the imaging data in the MIR-S channel, we found some datasets with two images in different observations. 
One image of each dataset accidentally showed a clear line of this scattered light caused by a bright point source illuminating the detector edge, while in the counterpart, the pointing direction was slightly dithered and the point source was properly imaged in the main field of view. 
We subtracted the counterpart image from the image with the scattered light after correcting for the position shift. 
This subtraction extracted the brightness due to the scattered light. 
We determined the brightness profile as a function of the distance from the edge of the detector. 
The intrinsic brightness, $S_{20}$, of the point source can be measured in the counterpart image. 
Figure~\ref{NPWscatter_prof_fig} shows the brightness profile, $S_{2}(X)$, of the scattered light normalized by $S_{20}$. 
%

\begin{figure}[!t]
\begin{center}
   \resizebox{0.85\hsize}{!}{
     \includegraphics[width=\linewidth]{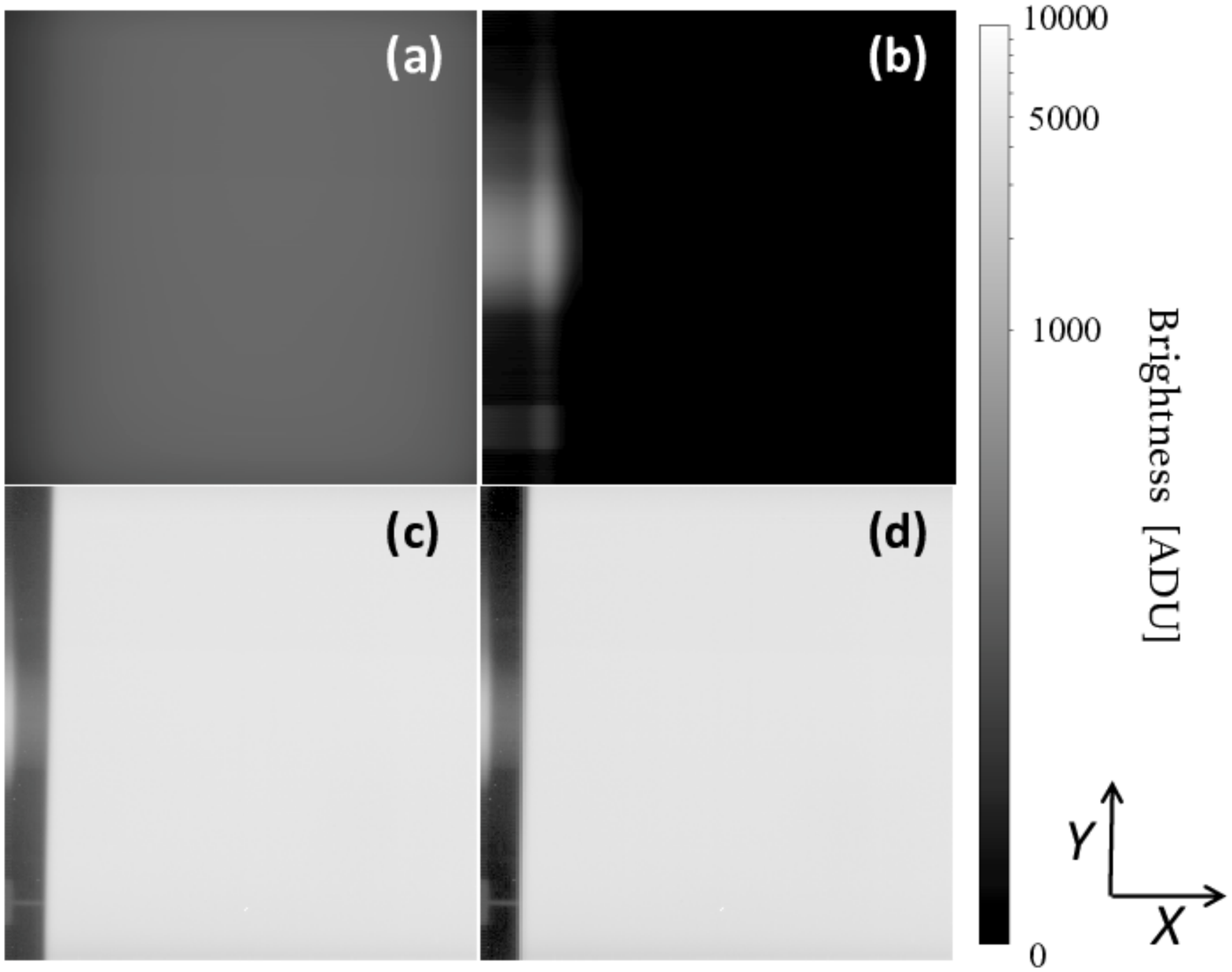}
   }
\end{center}
\caption{Two-dimensional images of (a) the light scattered in detector pixels from the main field of view scaled by a factor 20, (b) the light scattered at the edge of a detector after going through the small window aperture scaled by a factor of 100, and before and after the subtraction of these scattered light, (c) and (d), respectively. This is an example in the case of data pointed at ($\lambda_{s}$,$\beta_{s}$) = (174.13 deg, -1.00 deg) in observation ID of 1500701.1 and obtained with SG2.}
\label{artifact_fits_fig}
\end{figure}

According to data obtained from the NIR channel, which was pointed in the same direction as the MIR-S channel and for which the images covered the field of view of the small aperture window, no bright point source contaminated the field of view of the small window in any of the data we used. 
We therefore assumed the brightness to be uniform in the field of view of the small window. 
Fortunately, the spectrum of the incident light passing through the edge of the small aperture window illuminates only a few pixel columns at the edge of the detector sensitive area. 
We replaced the brightness in such pixel columns by the brightness illuminating the detector edge, $S_{20}(Y)$, and then calculated the brightness distribution of $S_{2}(X,Y)$ as in panel (b) of Figure~\ref{artifact_fits_fig}. 
Note that we optimized some parameters of the profile $S_{2}(X)$ described in Figure~\ref{NPWscatter_prof_fig} for the diffuse source of the scattered light. 

\subsection{The ghost of the small aperture window}

\begin{figure}[!t]
\begin{center}
     \includegraphics[width=\linewidth]{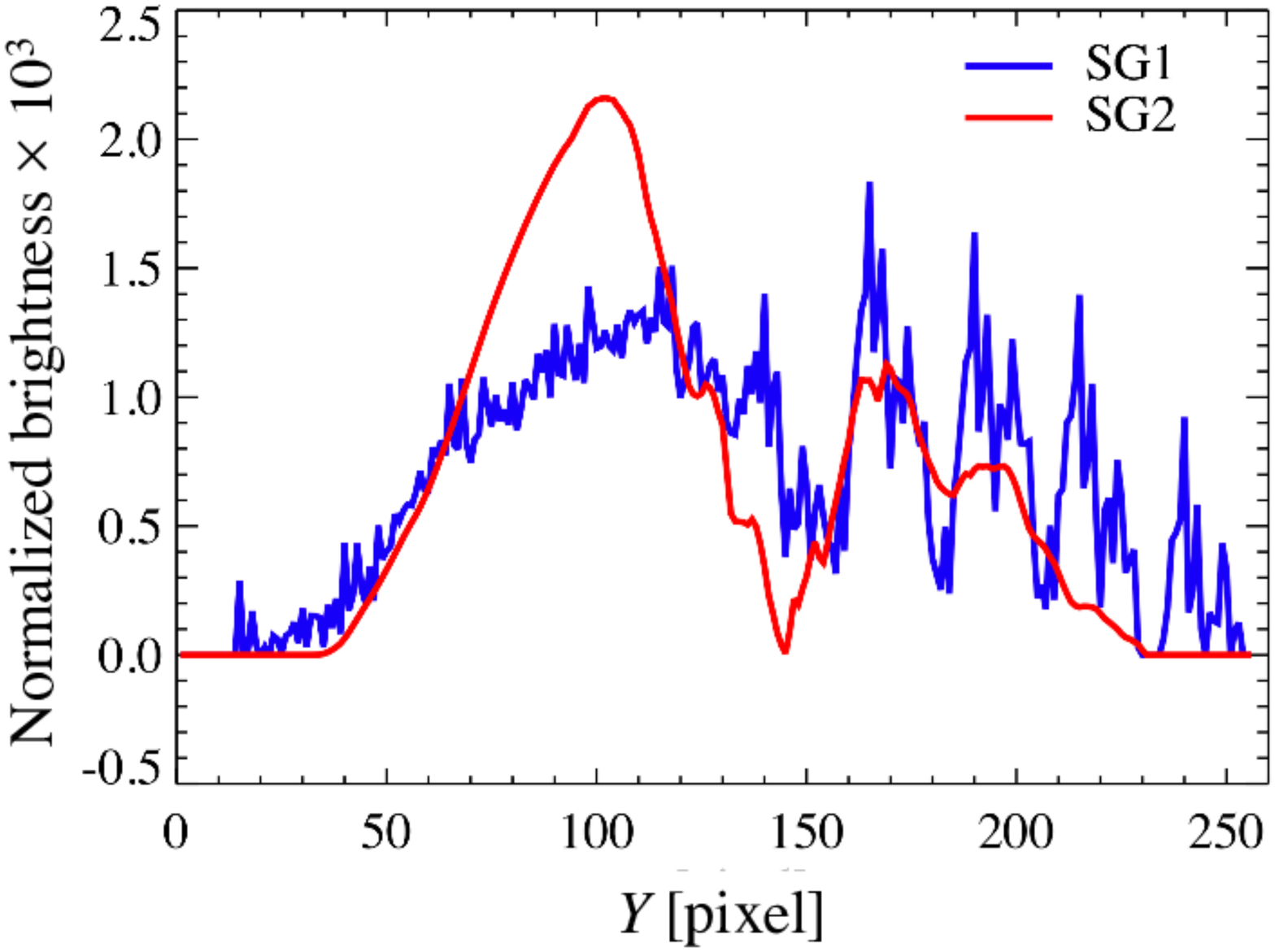}
\end{center}
\caption{Brightness profile of the ghost of the small aperture window, normalized by the typical brightness in the small aperture window. Here $Y$ is the direction of the wavelength dispersion, as defined in Figures \ref{artifact_schema_fig} and \ref{artifact_fits_fig}. This corresponds to the quantity $G_{3}(Y)$ in equation (\ref{ghost}).}
\label{ghost_prof_fig}
\end{figure}

Some fraction of the light that passes through the small aperture window makes ghost components in the slit-spectroscopic region after tracing a ray like (3) in Figure~\ref{artifact_schema_fig}. 
We assumed that this did not contaminate the spectroscopic region of the small aperture window itself. 
As mentioned above, spectra from the small aperture window are partly seen at the edge of the detector sensitive area in the spectroscopic images. 
By combining the spectra of the small aperture window and the slit, we estimated the brightness profile of the ghost by the following steps, after first subtracting the artifacts described in the previous two subsections. 
\begin{enumerate}
\item After extracting the one-dimensional spectra from the slit (slit spectra), we convolved them with the width of the small window (corresponding to 25 pixels, while the slit width is $\sim$ 2 pixels). Since the zodiacal emission is just a uniform background on this spatial scale, this convolved spectrum can be considered as the model spectrum of the small aperture window, including the convolved ghost. 
\item We extracted one-dimensional spectra of the small aperture window (window spectra) from the data at the edge columns of the pixel array and subtracted them from the model spectra derived in the previous step. Since the extracted window spectra are assumed not to include the ghost, this subtraction produces the brightness profile of the ghost convolved with the width of the small aperture window. 
\item We de-convolved the profile with the width of the small aperture window. 
\end{enumerate}
The shape of the derived profile seems to be common in all the data, and the absolute value is roughly proportional to the typical value of the mid-infrared brightness incident through the small aperture window. 
Figure~\ref{ghost_prof_fig} shows the normalized ghost profile obtained from the typical brightness in the small aperture window. 

\begin{figure}[!t]
\begin{center}
     \includegraphics[width=\linewidth]{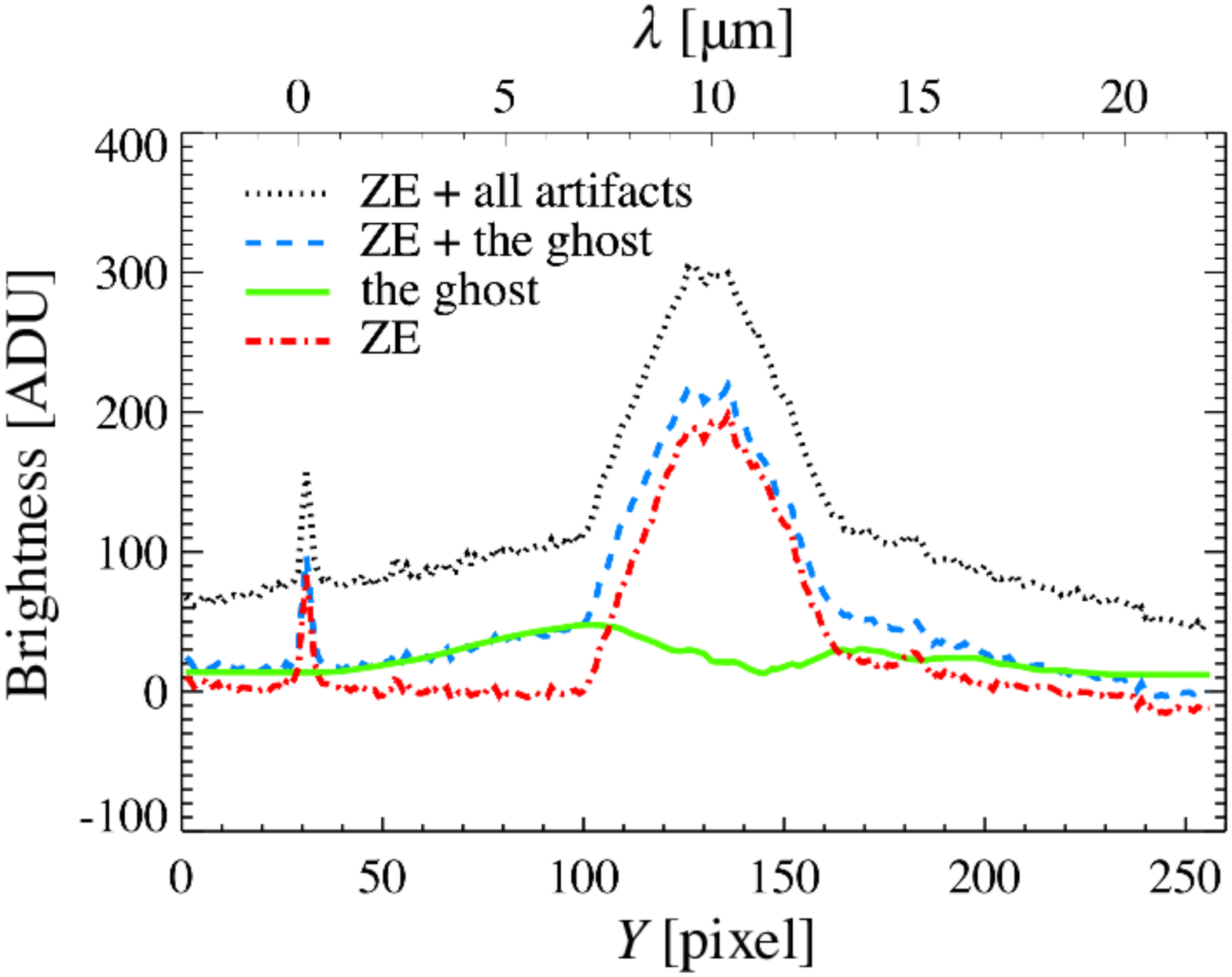}
\end{center}
\caption{Brightness profiles in the slit-spectroscopic region. The brightness originates from zodiacal emission, denoted by ZE, with all artifacts (black); with only the ghost component after subtracting the other two artifact components of the scattered lights (blue); and with no artifacts after subtracting all the artifacts (red). The green line represents the brightness profile of the ghost component and corresponds to the difference between the blue and red curves. The quantity $Y$ is the pixel position along the direction of spectral dispersion, as defined in Figures \ref{artifact_schema_fig} and \ref{artifact_fits_fig}. At the top, we show the corresponding wavelengths calculated using our wavelength calibration method. The 0th- and 2nd-order light is seen at $Y \sim 30$ and $160 \leq Y \leq 200$, respectively. This is an example of data obtained with SG2 during pointing in the direction ($\lambda_{s}$,$\beta_{s}$) = (174.13 deg, -1.00 deg). The observation ID is 1500701.1.}
\label{ADUprofiles_fig}
\end{figure}

Using this common profile, we determined the absolute value appropriate for each pointing dataset. 
Since we could not determine the typical brightness in the small aperture window directly from the image obtained in the MIR-S channel, we instead fitted the absolute value of the profile to the observed data. 
If the artifacts are completely removed, the brightness in the regions where the optical system has no response should be zero. 
However, the brightness in such regions was actually non-zero, even after subtracting the other two artifact components, (1) and (2). 
We assumed this non-zero brightness to originate from the ghost of the small aperture window. 
We therefore fitted the following function $S_{3}(Y)$ to the non-zero brightness seen in the regions without a system response:
\begin{equation}
S_{3}(Y) = a + b~Y + c~G_{3}(Y) \mbox{~~~[ADU]}, 
\label{ghost}
\end{equation}
where $G_{3}(Y)$ is the brightness profile shown in Figure~\ref{ghost_prof_fig}. 
The linear terms $a + b~Y$ represents the linear offset needed to correct the residual. 
This offset may correspond to the ghost component of the main field of view, although we could not conclude this definitively. 
The constants $a$, $b$, and $c$ are free parameters that we optimized individually for each pointing dataset. Thus, $S_{3}(Y)$ becomes identical to the non-zero brightness in the regions without a system response. 
As such regions, we selected the pixel ranges $25 \leq Y \leq 80$ or $180 \leq y \leq 240$ for SG1, and $45 \leq Y \leq 90$ or $200 \leq Y \leq 240$ for SG2, in order to avoid the contribution of the target brightness in the 0th, 1st, and 2nd dispersion order. 
Examples of the brightness profiles before and after subtracting the ghost component, and the profile of the ghost component itself, are shown in Figure~\ref{ADUprofiles_fig}. 

\begin{figure}[htbp]
\begin{center}
     \includegraphics[width=\linewidth]{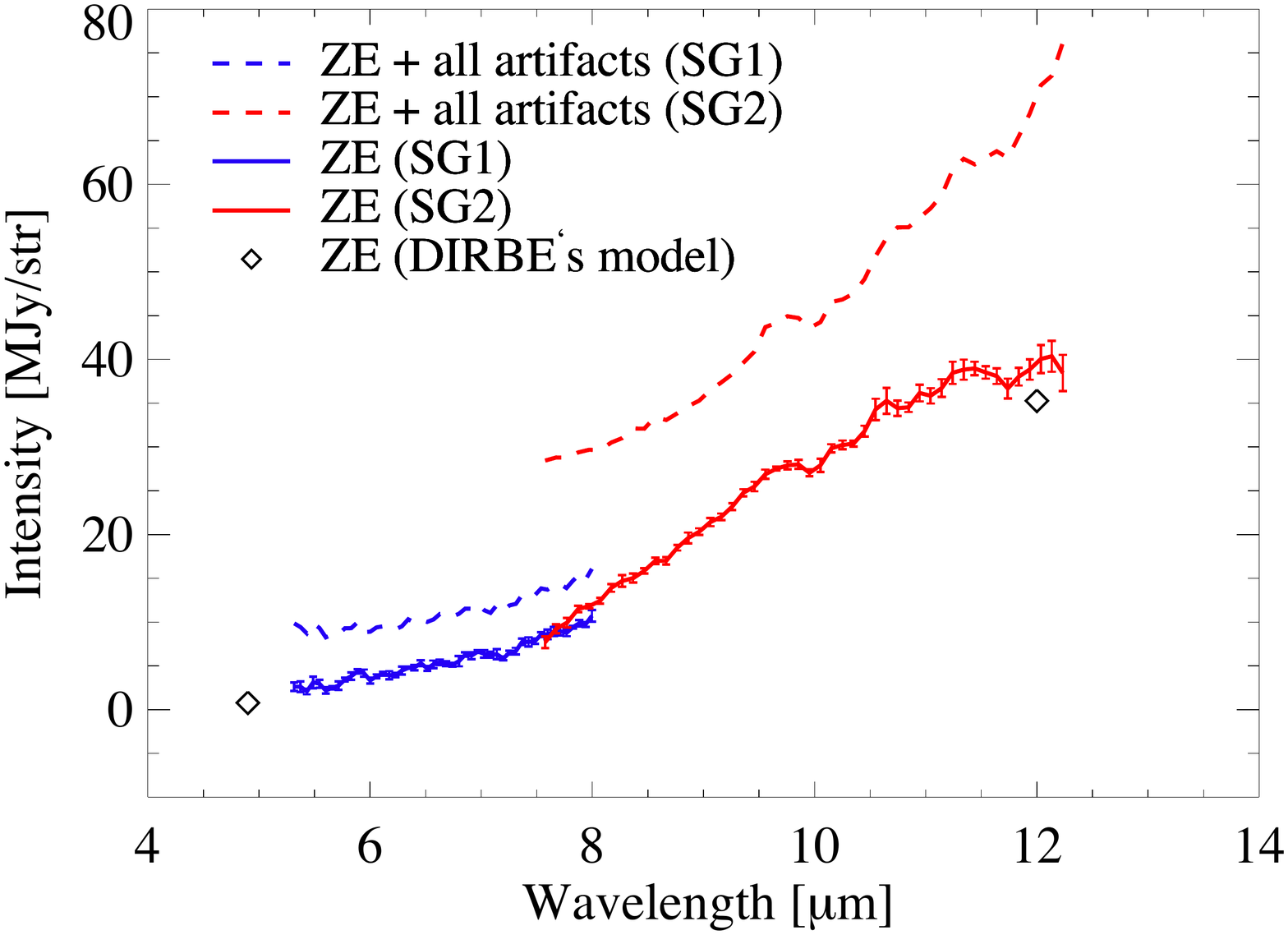}
\end{center}
\caption{Spectra obtained with SG1 and SG2 before and after subtracting the three types of artifacts. This is an example of data obtained while pointing in the direction ($\lambda_{s}$,$\beta_{s}$) = (174.13 deg, -1.00 deg). The observation ID is 1500701.1.}
\label{spec_compare_fig}
\end{figure}

After subtracting of these three types of artifacts, we succeeded in deriving spectra that are connected reasonably smoothly between the data from SG1 and SG2 for all pointing data. 
Figure~\ref{spec_compare_fig} shows an example of a spectrum obtained before and after the subtraction. 
It is clear that the subtraction of artifacts corrects for the inconsistency in the intensity levels of SG1 and SG2 in the overlapping wavelength range. 

\clearpage

\section{Obtained observed/continuum ratio at each direction}
\label{allemissivity}

We show observed/continuum ratios in all 74 sky directions from the next page. 
They are obtained by following the reduction flow described in section \ref{reduction_sec} and dividing by the continuum calculated from the equation (\ref{continuum_eq}). 

\begin{figure*}[htbp]
\begin{center}
     \includegraphics[height=1\linewidth, angle=90]{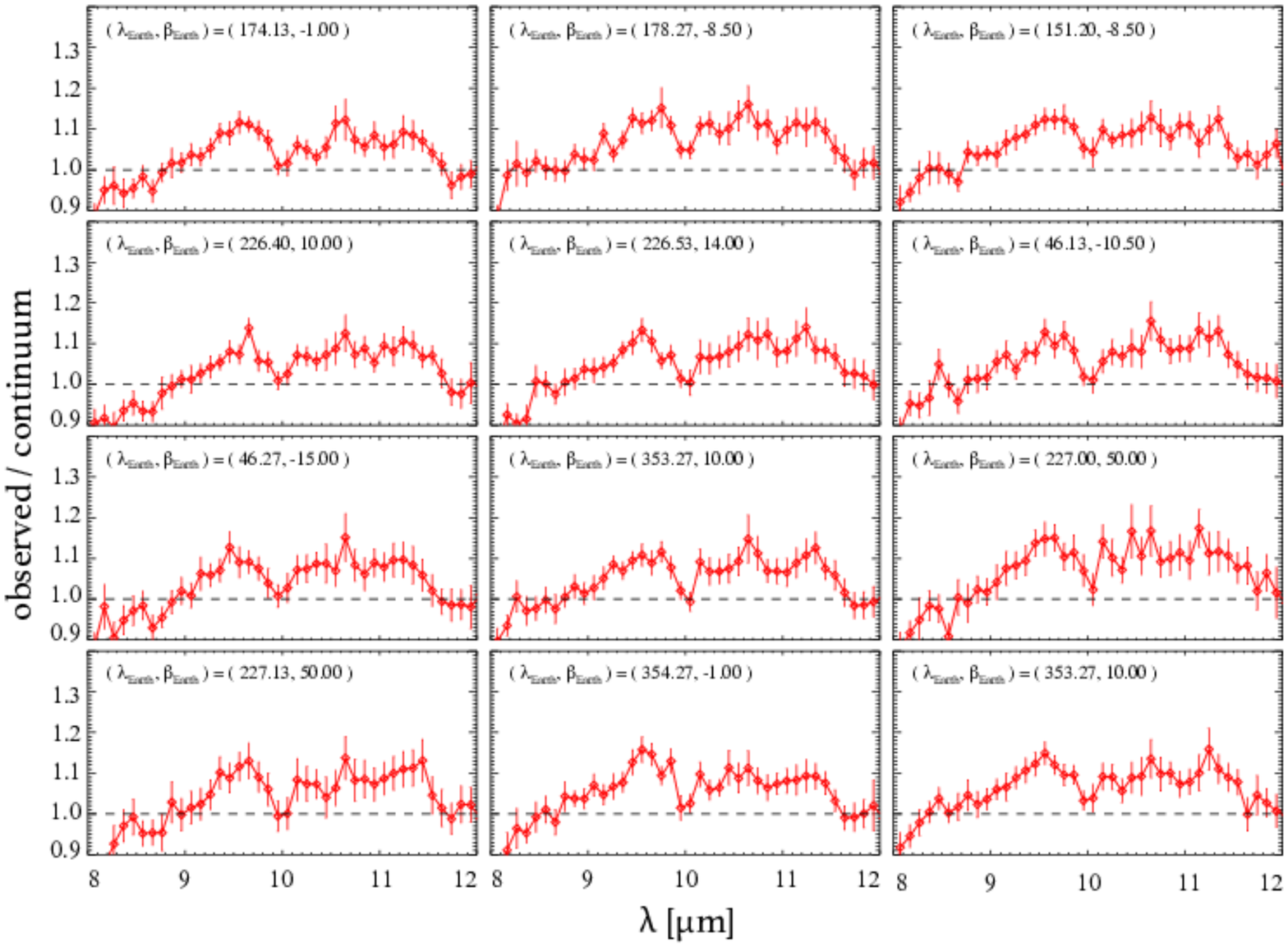}
\end{center}
\end{figure*}

\begin{figure*}[htbp]
\begin{center}
     \includegraphics[height=1\linewidth, angle=90]{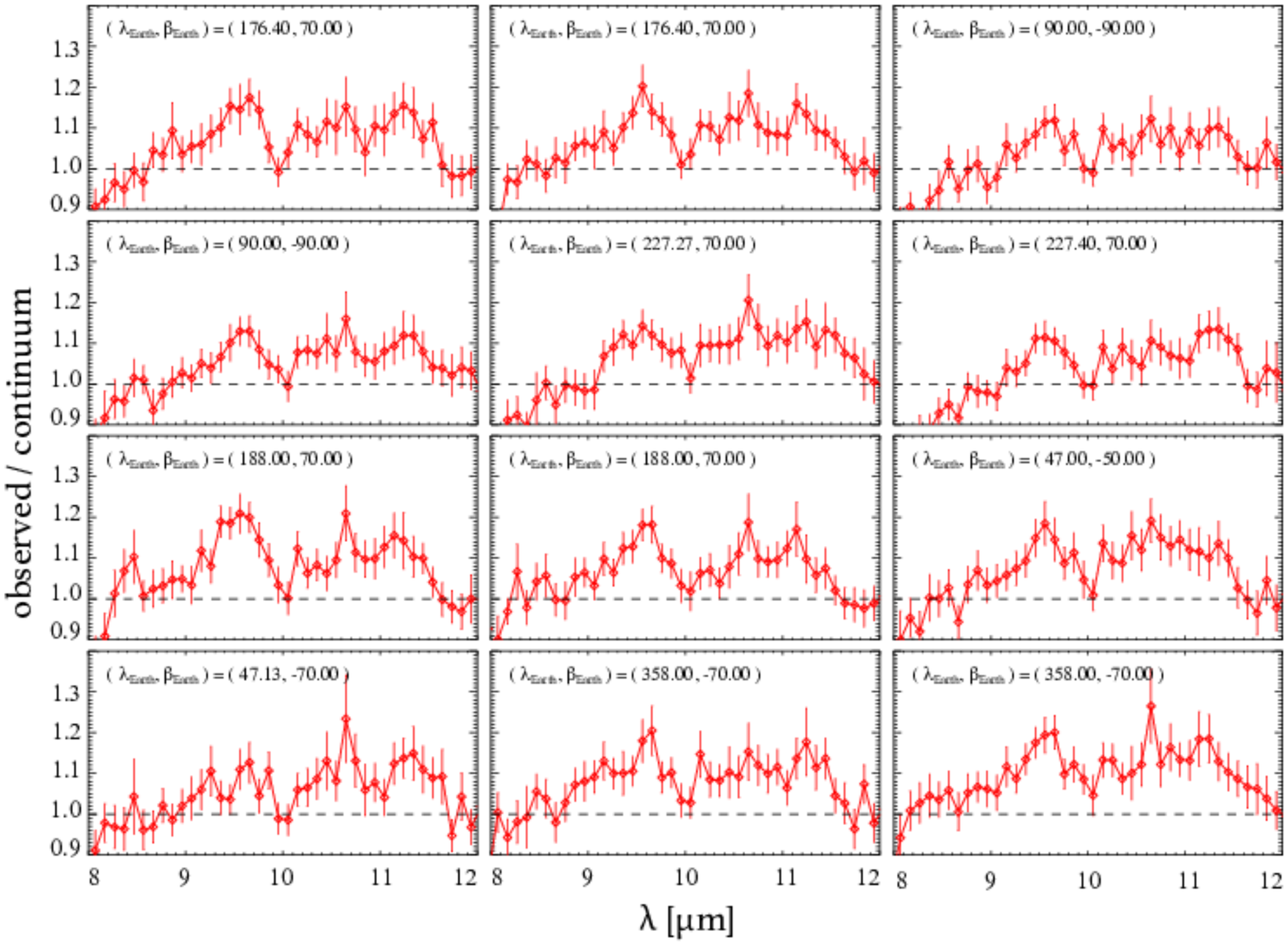}
\end{center}
\end{figure*}

\begin{figure*}[htbp]
\begin{center}
     \includegraphics[height=1\linewidth, angle=90]{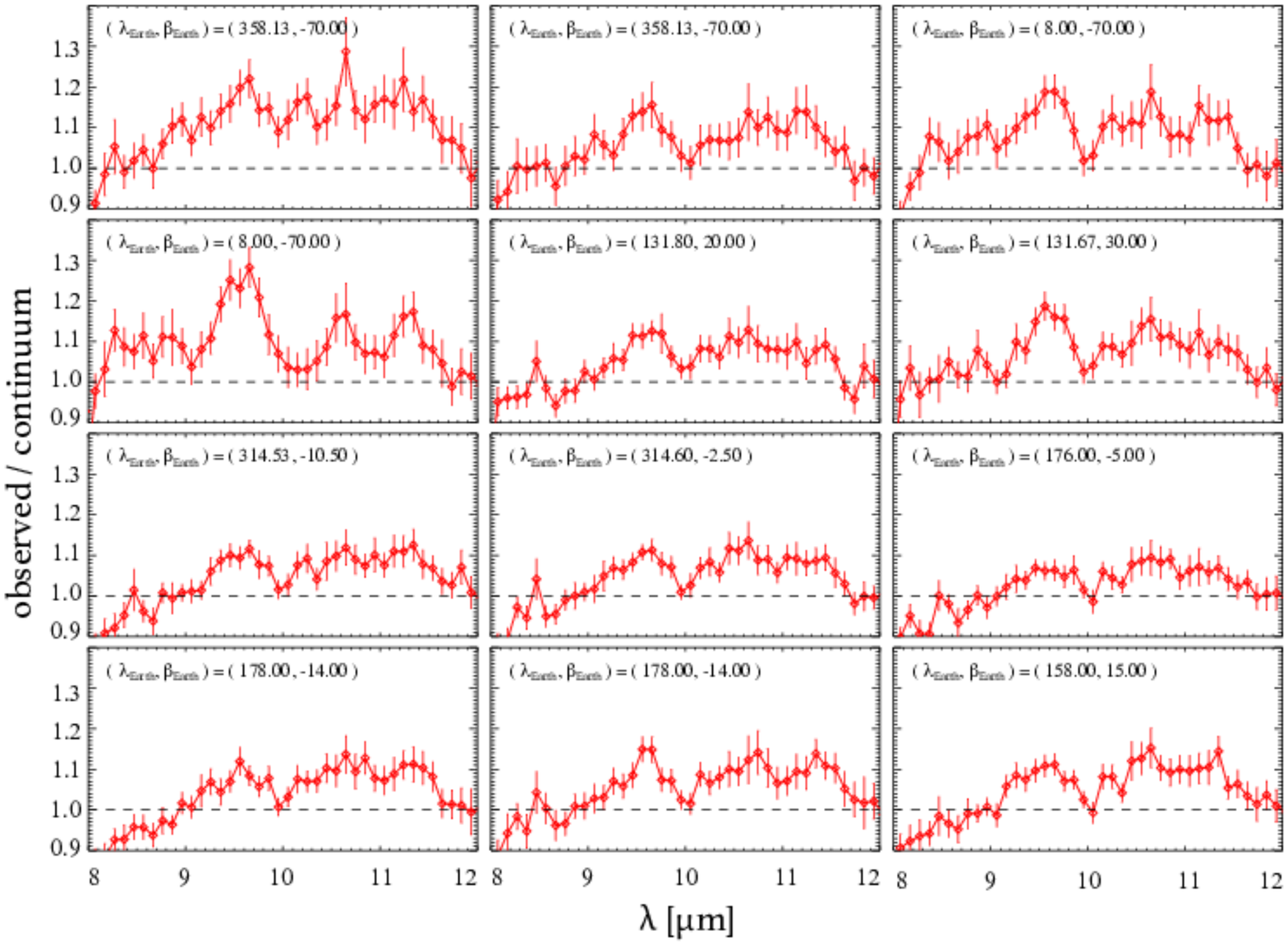}
\end{center}
\end{figure*}

\begin{figure*}[htbp]
\begin{center}
     \includegraphics[height=1\linewidth, angle=90]{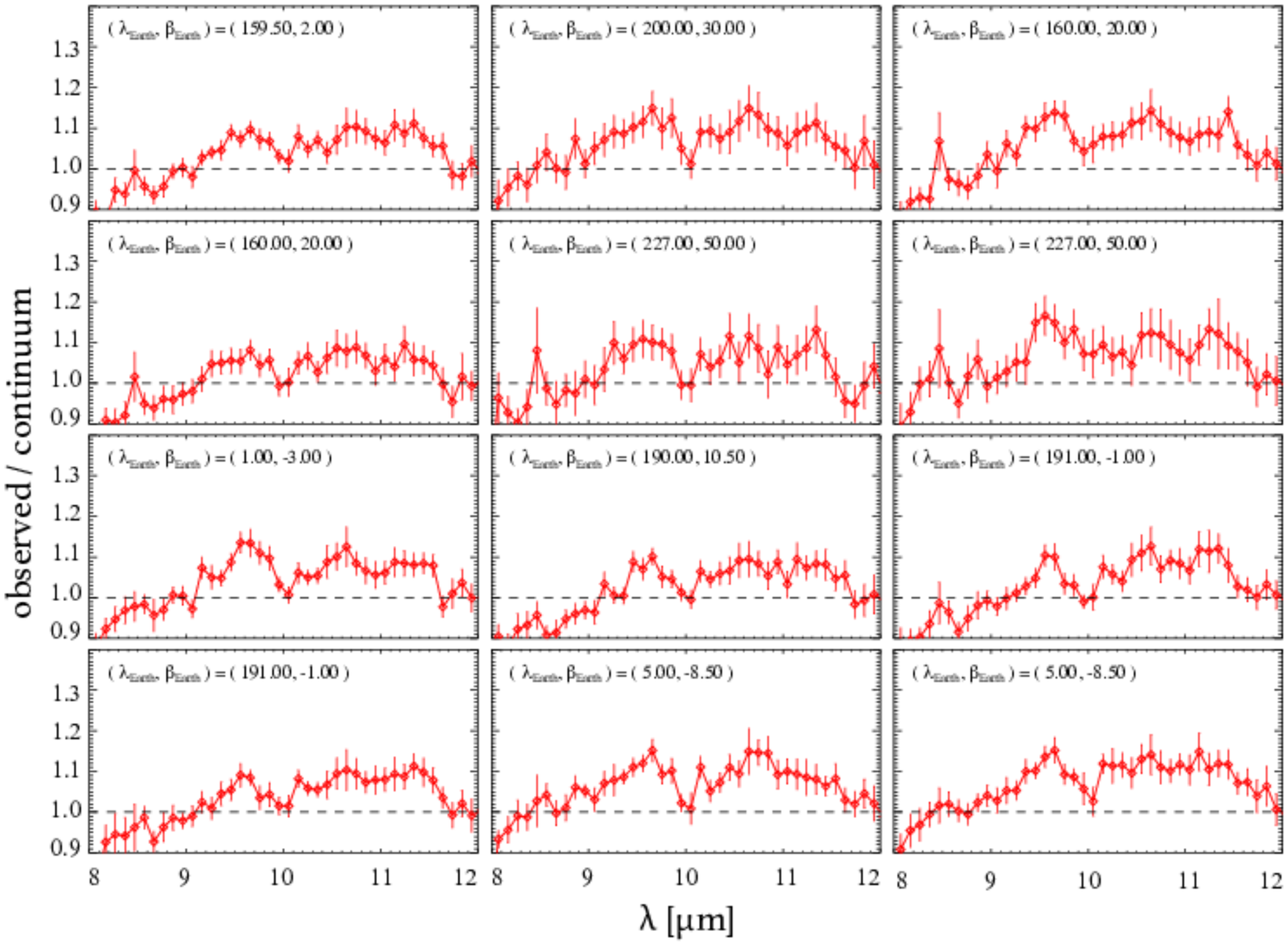}
\end{center}
\end{figure*}

\begin{figure*}[htbp]
\begin{center}
     \includegraphics[height=1\linewidth, angle=90]{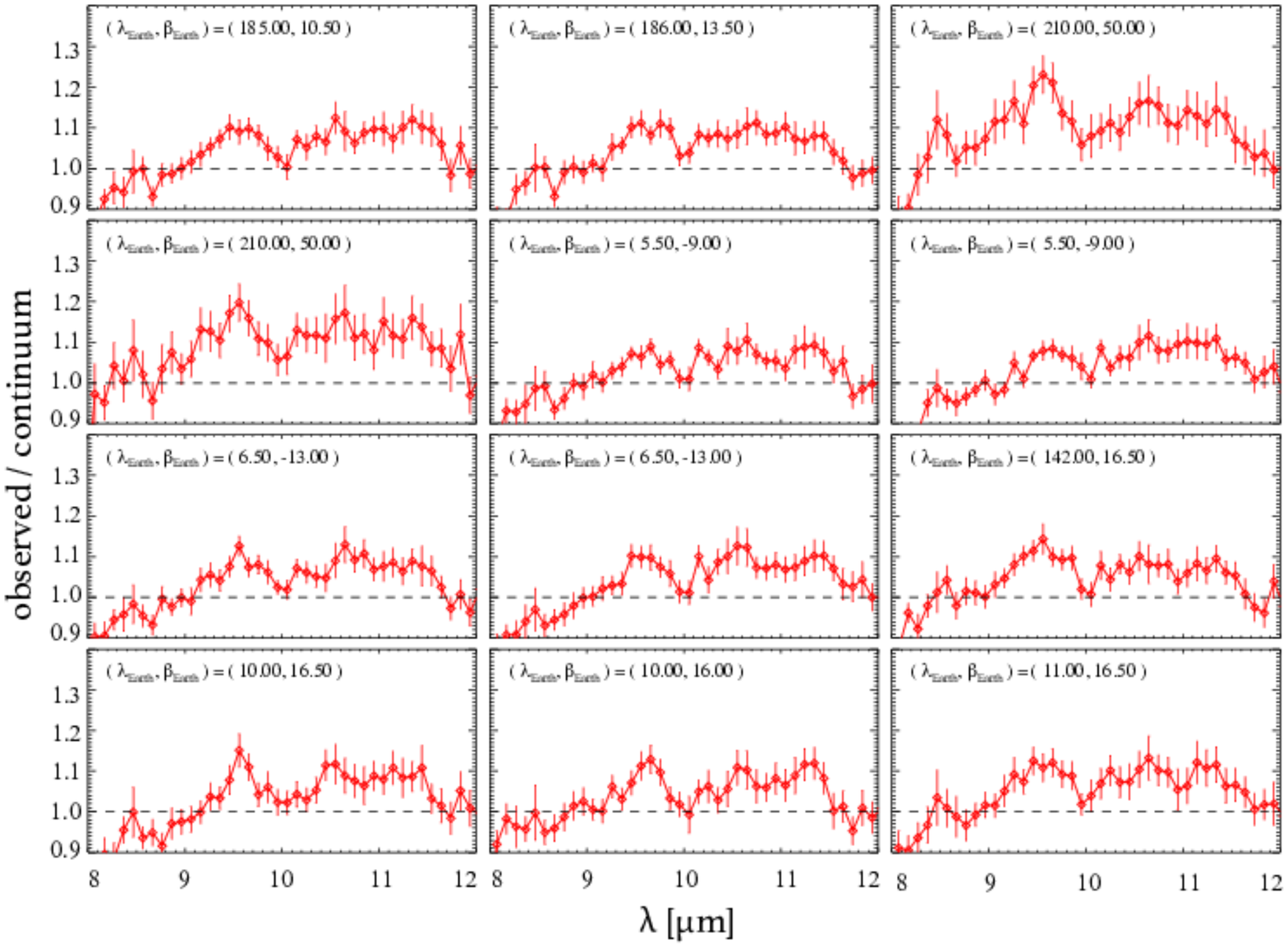}
\end{center}
\end{figure*}

\begin{figure*}[htbp]
\begin{center}
     \includegraphics[height=1\linewidth, angle=90]{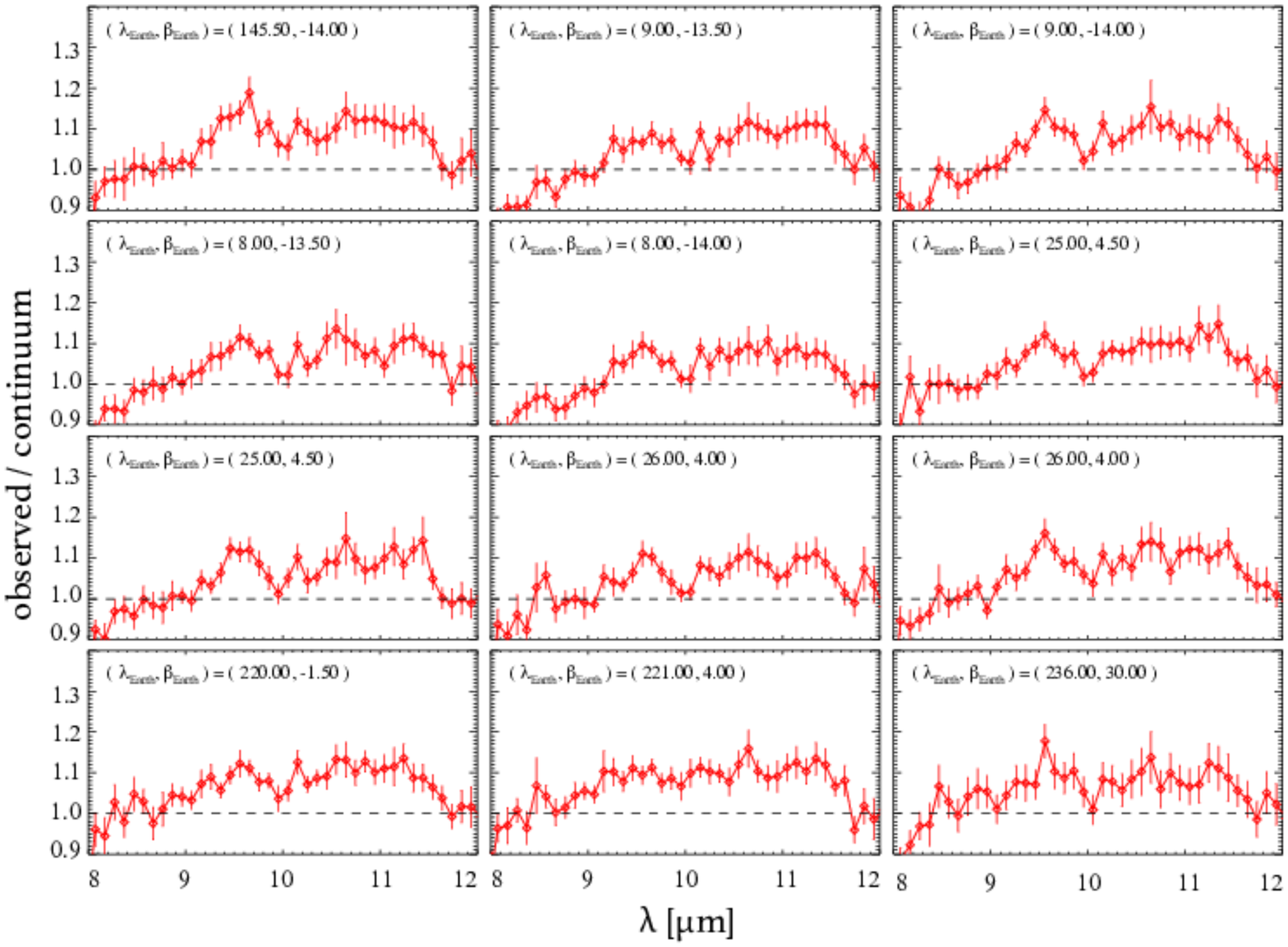}
\end{center}
\end{figure*}

\begin{figure*}[htbp]
\begin{center}
     \includegraphics[height=1\linewidth, angle=90]{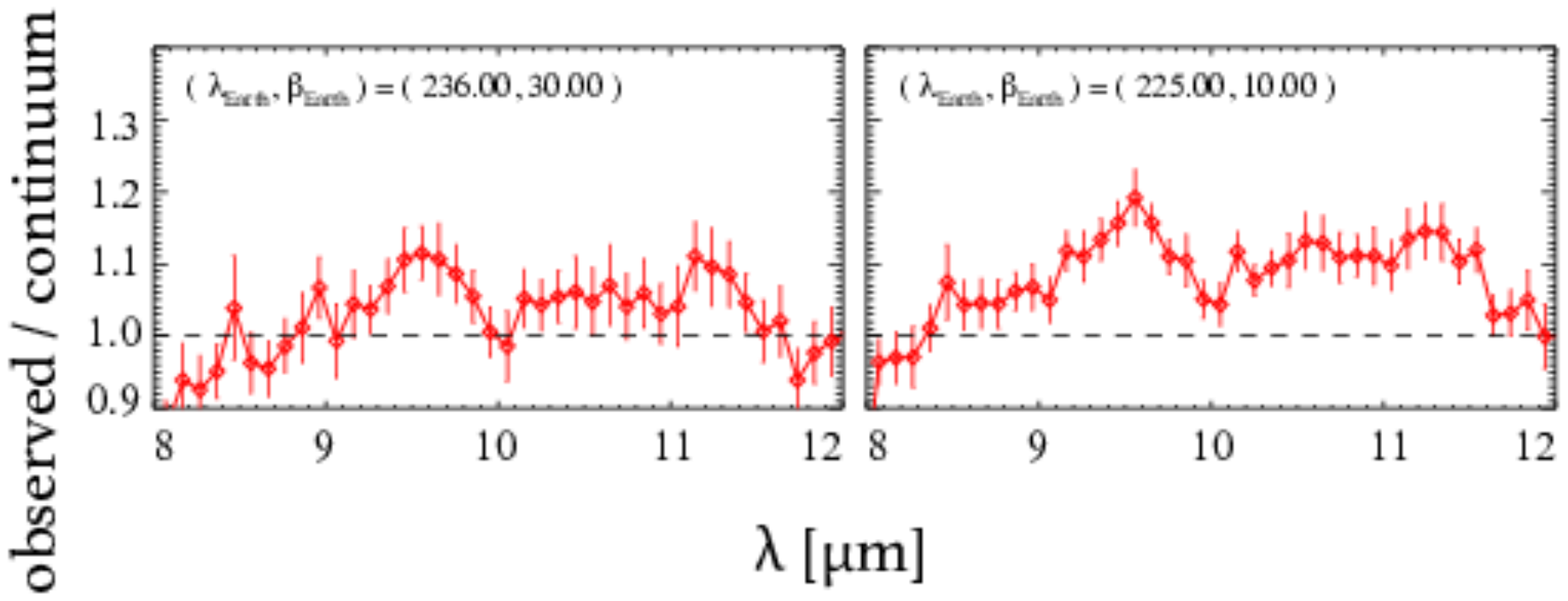}
\end{center}
\end{figure*}


\clearpage

\begin{thebibliography}{99}

\bibitem{Barker_1973} Barker, A.~S. \ 1973, Physical Review B, 7, 2507
\bibitem{Beck_2014} Beck, P., Garenne, A., Quirico, E., et al. \ 2014, Icarus, 229, 263
\bibitem{Bell_1997} Bell, K.~R., Cassen, P.~M., Klahr, H.~H., \& Henning, T. \ 1997, ApJ, 486, 372
\bibitem{Bernard-Salas_2003} Bernard-Salas, J., Pottasch, S.~R., Wesselius, P.~R., \& Feibelman, W.~A. \ 2003, A\&A, 406, 165
\bibitem{Bischoff_1998} Bischoff, A. \ 1998, Meteoritics and Planetary Science, 33, 1113
\bibitem{Bohren_1977} Bohren, C.~F., \& Wickramasinghe, N.~C. \ 1977, Ap\&SS, 50, 461
\bibitem{Bohren_1983} Bohren, C.~F., \& Huffman, D.~R. \ 1983, Absorption and Scattering of Light by Small Particles
\bibitem{Bradley_1983} Bradley, J.~P., Brownlee, D.~E., \& Veblen, D.~R. \ 1983, Nature, 301, 10
\bibitem{Bradley_1999} Bradley, J.~P., Keller, L.~P., Snow, T.~P., et al. \ 1999, Science, 285, 1716
\bibitem{Burns_1979} Burns, J.~A., Lamy, P.~L., \& Soter, S. \ 1979, Icarus, 40, 1
\bibitem{Chihara_2002} Chihara, H., Koike, C., Tsuchiyama, A., et al. \ 2002, A\&A, 391, 267
\bibitem{Cohen_1995} Cohen, M., Witteborn, F.~C., Walker, R.~G., Bregman, J.~D., \& Wooden, D.~H. \ 1995, AJ, 110, 275
\bibitem{Cohen_1996} Cohen, M., Witteborn, F.~C., Carbon, D.~F., et al. \ 1996, AJ, 112, 2274
\bibitem{Cohen_1999} Cohen, M., Walker, R.~G., Carter, B., et al. \ 1999, AJ, 117, 1864
\bibitem{Cohen_2003} Cohen, M., Wheaton, W.~A., \& Megeath, S.~T. \ 2003, AJ, 126, 1090
\bibitem{Demichelis_2012} Demichelis, R., Suto, H., N\"{o}el, Y., et al. \ 2012, MNRAS, 420, 147
\bibitem{Dorschner_1995} Dorschner, J., Begemann, B., Henning, Th., J\"{a}ger, C., \& Mutschke, H. \ 1995, A\&A, 300, 503
\bibitem{Gail_2001} Gail, H.~P. \ 2001, A\&A, 378, 192
\bibitem{Gail_2004} Gail, H.~P. \ 2004, A\&A, 413, 571
\bibitem{Ganguly_1995} Ganguly, J., \& Bose, K. \ 1995, in Lunar and Planetary Inst. Technical Report, 26, Lunar and Planetary Science Conference
\bibitem{Genzel_1973} Genzel, L., \& Martin, T.~P.  \ 1973, Surface Science, 34, 33
\bibitem{Germani_1990} Germani, M.~S., Bradley, J.~P., \& Brownlee, D.~E. \ 1990, Earth and Planetary Science Letters, 101, 162
\bibitem{Grun_1985} Gr\"{u}n, E., Zook, H.~A., Fechtig, H., \& Giese, R.~H. \ 1985, Icarus, 62, 244
\bibitem{Hallenbeck_2000} Hallenbeck, S.~L., Nuth III, J.~A., \& Nelson, R.~N. \ 2000, ApJ, 535, 247
\bibitem{Jehn_2000} Jehn, R.\ 2000, P\&SS, 48, 1429
\bibitem{Keller_2005} Keller, L.~P., \& Messenger, S. \ 2005, in Astronomical Society of the Pacific Conference Series, 341, Chondrites and the Protoplanetary Disk, 657
\bibitem{Kelsall_1998} Kelsall, T., Weiland, J.~L., Franz, B.~A., et al. \ 1998, ApJ, 508, 44
\bibitem{Kemper_2005} Kemper, F., Vriend, W.~J., \& Tielens, A.~G.~G.~M. \ 2005, ApJ, 633, 534
\bibitem{Kerker_1969} Kerker, M. \ 1969, The scattering of light and other electromagnetic radiation. 
\bibitem{Kimura_2013} Kimura, H. \ 2013, ApJ, 775, 18
\bibitem{Koike_2003} Koike, C., Chihara, H., Tsuchiyama, A., et al. \ 2003, A\&A, 399, 1101
\bibitem{Leinert_2002} Leinert, C., \'{A}brah\'{a}m, P., Acosta-Pulido, J., Lemke, D., \& Siebenmorgen, R. \ 2002, A\&A, 393, 1073
\bibitem{Liou_1995} Liou, J.~C., Dermott, S.~F., \& Xu, Y.~L. \ 1995, Planet. Space Sci., 43, 717
\bibitem{McSween_1979} McSween, H.~Y. \ 1979, Reviews of Geophysics and Space Physics, 17, 1059
\bibitem{McSween_1987} McSween, H.~Y. \ 1987, Geochimica et Cosmochimica Acta, 51, 2469
\bibitem{Merouane_2014} Merouane, S., Djouadi, Z., \& Le Sergeant d'Hendecourt, L. \ 2014, ApJ, 780, 174
\bibitem{Messenger_2013} Messenger, S., Keller, L.~P., \& Nguyen, A.~N. \ 2013, Proceedings of Science, LCDU 2013, 40
\bibitem{Mosteller_1977} Mosteller, F., \& Tukey, J.~W. \ 1977, Data analysis and regression. A second course in statistics
\bibitem{Murakami_2007} Murakami, H., Baba, H., Barthel, P., et al. \ 2007, PASJ, 59, S369
\bibitem{Murata_2009} Murata, K., Chihara, H., Koike, C., et al. \ 2009, ApJ, 697, 836
\bibitem{Nesvorny_2003} Nesvorn\'{y}, D., Bottke, W.~F., Levison, H.~F., \& Dones, L. \ 2003, ApJ, 591, 486
\bibitem{Nesvorny_2006} Nesvorn\'{y}, D., Vokrouhlick\'{y}, D., Bottke, W.~F., \& Sykes, M.~V. \ 2003, ApJ, 591, 486
\bibitem{Nesvorny_2010} Nesvorn\'{y}, D., Jenniskens, P., Levison, H.~F., Bottke, W.~F., Vokrouhlick\'{y}, D., \& Gounelle, M. \ 2010, ApJ, 713, 816
\bibitem{Noguchi_2011} Noguchi, T., Nakamura, T., Kimura, M., et al. \ 2011, Science, 333, 1121
\bibitem{Ohyama_2007} Ohyama, Y., Onaka, T., Matsuhara, H., et al. \ 2007, PASJ, 59, S411
\bibitem{Onaka_2007} Onaka, T., Matsuhara, H., Wada, T., et al. \ 2007, PASJ, 59, S401
\bibitem{Ootsubo_1998} Ootsubo, T., Onaka, T., Yamamura, I., Tanab\'{e}, T., Roellig, T.~L., Chan, K.~W., \& Matsumoto, T. \ 1998, Earth Planets Space, 50, 507
\bibitem{Poppe_2016} Poppe, A.~R. \ 2016, Icarus, 264, 369
\bibitem{Reach_2003} Reach, W.~T., Morris, P., Boulanger, F., \& Okumura, K. \ 2003, Icarus, 164, 384
\bibitem{Sakon_2007} Sakon, I., Onaka, T., Wada, T., et al. \ 2007, PASJ, 59, S483
\bibitem{Sasaki_2001} Sasaki, S., Nakamura, K., Hamabe, Y., Kurahashi, E., \& Hiroi, T. \ 2001, Nature, 410, 555
\bibitem{Schramm_1989} Schramm, L.~S., Brownlee, D.~E., \& Wheelock, M.~M. \ 1989, Meteoritics, 24, 99
\bibitem{Shinnaka_2018} Shinnaka, Y., Ootsubo, T., Kawakita, H., et al. \ 2018, AJ, 156, 242
\bibitem{Sogawa_2006} Sogawa, H., Koike, C., Chihara, H., Suto, H., Tachibana, S., Tsuchiyama, A., \& Kozasa, T. \ 2006, A\&A, 451, 357
\bibitem{Strazzulla_2005} Strazzulla, G., Dotto, E., Binzel, R., et al. \ 2005, Icarus, 174, 31
\bibitem{Sykes_1986} Sykes, M.~V., \& Greenberg, R. \ 1986, Icarus, 65, 51
\bibitem{Takigawa_2012} Takigawa, A., \& Tachibana, S. \ 2012, ApJ, 750, 149
\bibitem{Tomeoka_1989} Tomeoka, K., McSween, Harry Y.~J., \& Buseck, P.~R. \ 1989, Antarctic Meteorite Research, 2, 221
\bibitem{Tukey_1977} Tukey, J.~W. \ 1977, Exploratory data analysis
\bibitem{Usui_2018} Usui, F., Hasegawa, S., Ootsubo, T., \& Onaka, T. \ 2019, PASJ, 71, 1
\bibitem{Vernazza_2009} Vernazza, P., Binzel, R.~P., Rossi, A., Fulchignoni, M., \& Birlan, M. \ 2009, Nature, 458, 993
\bibitem{Wyatt_1999} Wyatt, M.~C., Dermott, S.~F., Telesco, C.~M., et al. \ 1999 ApJ, 527, 918



\end{thebibliography}
\end{document}